\documentclass[suppldata]{interact}

\usepackage{epstopdf}% To incorporate .eps illustrations using PDFLaTeX, etc.
\usepackage[caption=false]{subfig}% Support for small, `sub' figures and tables

\usepackage{braket}
\usepackage{comment}
\usepackage{color}

\usepackage[numbers,sort&compress,merge]{natbib}% Citation support using natbib.sty
\bibpunct[, ]{[}{]}{,}{n}{,}{,}% Citation support using natbib.sty
\theoremstyle{plain}% Theorem-like structures provided by amsthm.sty

\theoremstyle{definition}

\theoremstyle{remark}

\begin{document}

%\articletype{ARTICLE TEMPLATE}% Specify the article type or omit as appropriate

\title{Low-rank decomposition on the antisymmetric product of geminals for strongly correlated electrons}

\author{
\name{Airi Kawasaki\textsuperscript{a}\thanks{CONTACT Airi Kawasaki. Email: airi.geminal@gmail.com} and Naoki Nakatani\textsuperscript{b}}
\affil{\textsuperscript{a}Division of Electronics and Informatics, Graduate school of Science and Technology, Gunma University, 1-5-1 Tenjin-cho, Kiryu-shi, Gunma 376-8515, Japan }
\affil{\textsuperscript{b}Department of Chemistry, Graduate School of Science, Tokyo Metropolitan University, 1-1 Minami-Osawa, Hachioji-shi, Tokyo 192-0397, Japan}
}

\maketitle

\begin{abstract}
We investigated some variational methods to compute a wavefunction based on antisymmetric product of geminals (APG).
The Waring decomposition on the APG wavefunction leads a finite sum of antisymmetrized geminal power (AGP) wavefunctions, each for which the variational principle can be applied.
We call this as AGP-CI method which provides a variational solution of the APG wavefunction efficiently.
However, number of AGP wavefunctions in the exact AGP-CI formalism become exponentially large in case of many-electron systems.
Therefore, we also investigate the low-rank APG wavefunction, in which the geminal matrices are factorized by the Schur decomposition.
Interestingly, only a few non-zero eigenvalues (up to half number of electrons) were found from the Schur decomposition on the APG wavefunction.
We developed some methods to approximate the APG wavefunction by lowering the ranks of geminal matrices, and demonstrate their performance.
Our new geminal method based on the low-rank decomposition can drastically reduce the number of variational parameters, although there is no efficient algorithm so far, due to some mathematical complications.
\end{abstract}

\begin{keywords}
geminal; strongly correlated systems; Hubbard model; Schur decomposition
\end{keywords}

\section{Introduction}

Accurate computation of an electronic state in many-electron systems is one of the most important and fundamental problems in the field of computational materials science \cite{aszabo82:qchem, olsen_bible}. 
So far, various electronic structure calculation methods such as Hartree-Fock (HF) theory, post-HF theories, and density-functional theory (DFT) \cite{1964PhRv..136..864H, kohn1965self} have been developed. However, methods such as HF and DFT, which are based on a one-body approximation, have the problem that they cannot incorporate interactions between electrons well. 
Interaction between electrons that is not explicitly taken into account in the HF method, is called electron correlation, and accurate description of electron correlation is particularly important to investigate strongly correlated electronic systems in case where the Coulomb force acting between electrons is strong.
For example, systems containing transition metal elements or rare-earth elements possess complicated electronic states due to strong correlation effects from d- or f-electrons. In such cases, accurate results cannot be obtained without sophisticated treatment of strongly correlated electrons, and therefore, the one-body approximation such as the HF method doesn't work anymore.
On the other hand, a set of excited configurations consisting of $N$-electron excitation from the HF wavefunction gives a $N$-body basis to compute the exact ground state.
This is called configuration-interaction (CI) method, and in case where all possible excited configurations for $N$-electron system are incorporated (called the full-CI), the exact ground state can be obtained within a given orbital subspaces, though its computational cost is too expensive to investigate systems containing a large number of electrons.
To avoid such complications, here we focus on the wavefunction spanned by up to the electron pair basis. 

A basis of the electron pair, also called by a geminal \cite{shull1959natural, 1965JMP.....6.1425C}, is an important concept to describing chemical bond in quantum chemistry and superconductivity in condensed-matter physics \cite{bardeen1957theory}.
Particularly in the context of geminals, there are many wavefunction theories had been developed.
Antisymmetric product of geminals (APG) \cite{kutzelnigg1964w, mcweeny1963density} is of the most basic formalisms, in which each geminal expresses a different chemical bond.

A geminal in the APG formalism can be defined as
\begin{eqnarray}
\hat{F}[k]\equiv \sum_{a,b}^{2K} F[k]_{ab} \hat{c}^{\dag}_a\hat{c}^{\dag}_b
\end{eqnarray}
where $F$ is a skew-symmetric matrix, $\hat{c}^{\dag}$ is the creation operator, $k$ represents the type of geminal, and $2K$ is the number of spin orbitals. Then, a total wavefunction in the APG formalism is of the form
\begin{eqnarray}
\ket{\Psi_{\mathrm{APG}}}= \hat{F}[1]\cdots\hat{F}[N/2]\ket{0}  \label{apg}
\end{eqnarray}
where $N$ is the number of electrons. 

Although the idea of the APG wavefunction has been around for a long time, there has not been well investigated to calculating the APG wavefunction as it is, due to the problem that the variational calculation of the APG is complicated and difficult \cite{limacher2013new}.
Therefore, various theories that approximate the APG wavefunction had been proposed for practical applications.

For example, a simplification called strong orthogonality is introduced to APG. The strong orthogonality condition imposes the following restrictions on geminal matrices:
\begin{eqnarray}
\sum_{c}^{2K} F[k]_{ca} F[l]_{cb} =0
\end{eqnarray}
for all $a$ and $b$. The APG wave function constructed from the strongly orthogonal geminals is called the antisymmetrized product of strongly orthogonal geminal (APSG) \cite{arai1960theorem, hurley1953molecular, Tokmachev:2016:PPG, tarumi2013accelerating, jeszenszki2015local, mcweeny1980molecular}. Recently, the development of perturbation theory based on APSG has also been reported \cite{jeszenszki2014perspectives}.

Another simplification is that each geminal is restricted to form a singlet pair by sharing the same spatial orbital for the up and down spins, which assumes that the unitary transformations of all geminal matrices are equal. The APG wavefunction under this condition is called the antisymmetrized product of interacting geminals (APIG) \cite{silver1969natural, johnson2017strategies, nicely1971geminal}. Recently, methods such as APr2G, which further simplifies APIG mathematically, have been developed \cite{johnson2013size, limacher2013new, johnson2017strategies}.

However, APSG and APIG are too simplified from the APG for the sake of mathematical convenience, to accurately evaluate the electron correlation effect.
Thus, before making approximations along the lines of mathematical convenience, it is necessary to study APG as it is in order to establish an approximation method that captures the essence of electron correlation. 

In this paper, we present a rank-dependent decomposition on a geminal matrix to approximate the APG wavefunction, and explore the essence of electron correlation by spectral analyses of eigenvalues in geminal matrices.
From the APG analysis, we clarified that the approximation that simplifies the wavefunction by lowering the rank of the geminal skew-symmetric matrix is effective.
First, we start from the variational solution of the antisymmetrized geminal power (AGP) wavefunction, by which the APG wavefunction is expanded by linear combinations of the AGP functions, {\it{so called}} the AGP-CI theory \cite{2015PhRvA..91f2504U}.
Second, we present the Schur decomposition which gives eigenvalues and eigenvectors of a skew-symmetric matrix, and apply it to analyze the eigenvalue spectrum of the APG wavefunction using illustrative systems such that the Hubbard model \cite{hubbard1963electron} and some small molecules.
Finally, we propose the rank-dependent approximations on the APG wavefunction and show some demonstrative applications and analyses to provide a new direction of the geminal-based wavefunction ansatz.

\section{Theory}

\subsection{Antisymmetrized Geminal Power (AGP) wavefunction}

Before looking into the variational calculation of APG, we first introduce a geminal wavefunction for which the variational formula already exists.
Unlike the APG wavefunction which is of the product of $N/2$ different geminals, a wavefunction consisting of the product of the $N/2$ identical geminals is known as the antisymmetrized geminal power (AGP) \cite{1965JMP.....6.1425C}.
\begin{eqnarray}
\ket{\Psi_{\mathrm{AGP}}} = \hat{F}^{\frac{N}{2}}\ket{0} \equiv \ket{F}
\end{eqnarray}
Note that an overlap between two different AGP wavefunctions and the Hamiltonian matrix elements, which are necessary for variational calculations, can be analytically determined \cite{onishi1966generator, 2012PhLB..715..219M}.
For example, an overlap between the AGP wavefunctions described by variational parameters $\lambda$ and $\mu$ is of the form,
\begin{eqnarray}
\braket{ F[\lambda]|F[\mu]}= \left.\exp \left(\frac{1}{2}\mathrm{tr}\left[ \ln (1+ F[\mu]F[\lambda]^{T}t)\right] \right)\right|_{t^{\frac{N}{2}}} \label{OYovl}.
\end{eqnarray}
Here $|_{t^{\frac{N}{2}}} $ means to extract the $N/2$ th-order coefficient of the polynomial with respect to the auxiliary variable $t$. See Appendix \ref{app1} for other formulas.

In our previous work \cite{kawasaki2018tensor}, we overcame the computational difficulties in the variational calculation of the APG wavefunction by using a tensor decomposition method developed in the field of mathematics to bring the APG wavefunction into the sum of AGP wavefunctions, as we discuss in the next subsection.

\subsection{Variational calculation of the APG wavefunction}

A decomposition of product of geminals into a linear combination of powers of polynomials can be carried out through the Waring decomposition \cite{oeding2013eigenvectors, comon1996decomposition, alexander1995polynomial}.
In mathematics, much effort has been made to find the minimum number of polynomials required to reconstruct the original product of geminals.
The simplest case of the Waring decomposition is of the Fischer's formula \cite{fischer1994sums} for the monomial of geminals, as follows:
\begin{eqnarray}
&&\hat{F}[1]\hat{F}[2]\cdots\hat{F}[n] \nonumber \\ 
&&= \frac{1}{2^{n-1}n!}\sum_{i_2=0}^1\cdots\sum_{i_n=0}^1 (-1)^{i_2 + \cdots + i_n}\left(\hat{F}[1]+(-1)^{i_2}\hat{F}[2]+\cdots + (-1)^{i_n}\hat{F}[n]\right)^n  \label{fischer}
\end{eqnarray}
For example, when $n=3$, the Fischer's formula gives:
\begin{eqnarray}
\hat{F}[1]\hat{F}[2]\hat{F}[3] 
&=& \frac{1}{24}\left[ (\hat{F}[1]+\hat{F}[2]+\hat{F}[3] )^3 -(\hat{F}[1]+\hat{F}[2]-\hat{F}[3] )^3 \right. \nonumber \\
 && \left. -(\hat{F}[1]-\hat{F}[2]+\hat{F}[3] )^3+(\hat{F}[1]-\hat{F}[2]-\hat{F}[3] )^3 \right] 
\end{eqnarray}
Consequently, by applying the Fischer's formula, the APG wavefunction can be transformed to a finite sum of different AGP wavefunctions (called by AGP-CI), as follows:.
\begin{eqnarray}
\ket{\Psi_{\mathrm{AGP-CI}}} = ¥
\sum_k^{kn}\hat{F}[k]^{\frac{N}{2}}\ket{0} 
\end{eqnarray}
Here $kn$ is the number of types of AGP wavefunctions that we considered. Using this formalism, we can apply the Eq (\ref{OYovl}) to each AGP wavefunction to perform the variational optimization of the APG wavefunction.
If we take $kn$ to be sufficiently large, the AGP-CI wavefunction becomes equivalent to the full-CI wavefunction which is better than the APG wavefunction.
In this study, however, for a fair comparison between APG and AGP-CI, we only considered $kn=N/2$ kinds of AGP's in the AGP-CI, so that the total numbers of variational parameters are the same as the APG.

\subsection{Low-rank decomposition of geminals}

The original form of the APG wavefunction is parameterized by $N/2$ different $2K\times 2K$ skew-symmetric matrices, and as a result, generated geminals are highly redundant to each other because all geminals are spanned by the same orbital subspace.
This complication can be avoided by forcing geminals to be strongly-orthogonalized (APSG as we introduced already).
In the APSG wavefunction, $K$ orthogonal orbitals are divided into $N/2$ parts, each of which is used to generate a small geminal. In other words, each geminal is always defined from a different orbital subspace and thus, all the geminals are always orthogonalized as far as orbitals are orthogonalized.
While the APSG wavefunction gives us a simple solution to obtain the variational ansatz of the approximated APG wavefunction, the result highly depends on the division of the orthogonal orbital subspaces.
To overcome this problem, we introduce here, the low-rank decomposition of geminals to obtain an efficient and natural approximation of the APG wavefunction.

Now, let us consider the Schur decomposition on each skew-symmetric matrices $\{F\}$ in the APG wavefunction.
The Schur decomposition transforms a $2K\times 2K$ skew-symmetric matrix into a band-diagonalized matrix through unitary rotation as,
\begin{eqnarray}
F=USU^* , \label{usu}
\end{eqnarray}
where $U$ is a $2K \times 2K$ unitary matrix and $S$ is
\begin{eqnarray}
S = \begin{pmatrix}
0 & \lambda^{(1)} &  &  & 0 \\
-\lambda^{(1)} & 0 & \ddots & &  \\
 & \ddots & \ddots & & \\
& & & 0 & \lambda^{(K)} \\
0 & & & -\lambda^{(K)} & 0
\end{pmatrix} \label{diag}.
\end{eqnarray}
Here, we obtain $K$ eigenvalues $\lambda$. Note that the eigenvalue of a skew-symmetric matrix is imaginary, that is given by $\lambda i$.
Our idea of the low-rank decomposition of geminals is to extract only a few eigenvalues which describe important parts of electron correlations and reconstruct the APG wavefunction.
This approximation that sets the other eigenvalues to be 0, lowers the rank of the geminal matrix.
Thus, we would like to call this method as the low-rank APG approximation in the latter subsections and we define one that extracts only one eigenvalue for each geminal as rank-1 APG, one that extracts only two eigenvalues as rank-2 APG, and so on so forth.

In the rank-1 APG, for example, the $k$-th geminal operator can be approximated as
\begin{eqnarray}
\hat{F}[k] &=& \sum_{ab}^{2K}F[k]_{ab} \hat{c}^{\dag}_{a}\hat{c}^{\dag}_{b}\nonumber  \\
 &\rightarrow& \sum_{ab}^{2K}\frac{\lambda[k]^{(1)}}{2}\left(U[k]_{a1}U^{*}[k]_{\bar{1}b} - U[k]_{a\bar{1}}U^{*}[k]_{1b} \right) \hat{c}^{\dag}_{a}\hat{c}^{\dag}_{b} \label{rank-1x}
\end{eqnarray}
where we defined the $2K$ spin orbital index by $\{1,\bar{1},2,\bar{2},\ldots,K,\bar{K}\}$.
Note that all the unitary rotations are different from each other, meaning that the eigenvectors from different geminals are not orthogonalized.

In this study, we consider that all the unitary rotations are identical ($U[1] = U[2] = \cdots = U[k]$) for the sake of simplicity.
This introduces a new APG-based representation of a many-body wavefunction parameterized by a small number of $\{\lambda[k]\}$ and the orbital rotations by $U$. 
On the other hand, the low-rank APG where the unitary rotations are not identical for each geminal is briefly discussed in Appendix \ref{app2}.
Although this leads a more flexible form of the low-rank APG, it is so far, practically useless due to some unavoidable complications.

\subsubsection{Low-rank APG[$\lambda, X$] }

To find the optimal unitary rotation with respect to the variational condition for molecular orbitals, the unitary matrix is parameterized by a skew-symmetric matrix $X$ and is given by the exponentiated form of it.
\begin{eqnarray}
U = \exp(X) \label{orbopt}
\end{eqnarray}
We call this low-rank APG as APG[$\lambda, X$] method, where the $\lambda$ and the skew-symmetric matrix $X$ are variational parameters.
For example, at rank-1, a geminal can be converted from Eq (\ref{rank-1x}) to rewrite creation operators with the optimal orbital bases
\begin{eqnarray}
\hat{F}[k] = \lambda[k]^{(k)} \hat{a}^{\dag}_k \hat{a}^{\dag}_{\bar{k}}
\end{eqnarray}
where the new creation operators are given by
\begin{eqnarray}
\sum_{a}^{2K} U_{ak} \hat{c}^{\dag}_{a} \equiv \hat{a}^{\dag}_{k} .
\end{eqnarray}
Consequently, the rank-1 APG wavefunction is equivalent to the HF wavefunction
\begin{eqnarray}
\ket{\Psi_{\mathrm{r1APG}} } = \prod_{k}^{N/2} \hat{F}[k] \ket{0}= \prod_{k}^{N/2} \lambda[k]^{(k)} \hat{a}^{\dag}_k \hat{a}^{\dag}_{\bar{k}} \ket{0}
\end{eqnarray}
where $\prod_{k}^{N/2} \lambda^{(k)} = 1$ for normalization condition.
Therefore, electron correlation effect appears from the rank-2 APG wavefunction.
In the rank-2 or higher-rank wavefunction, however, there are many possibility to represent the low-rank APG wavefunction due to the choice of orbital subspace to generate low-rank geminals.
For example, if the orbital subspace of each geminal is independent from the others, it leads strongly-orthogonalized variant of the low-rank APG.
On the other hand, if some orbital subspaces are shared with the different geminals, these geminals are not orthogonalized and interacting with each other.
Detailed examples of the low-rank APG[$\lambda,X$] considered in this study are provided in \ref{low-rank_x}.

Variational calculation with the APG[$\lambda,X$] is carried out by optimizing $\lambda$ and $U$ separately.
First, we compute the wavefunction to minimize the energy with respect to a certain formalism of the low-rank APG parameterized by $\lambda$.
Next, one- and two-body integrals are transformed by the unitary rotation $U$ obtained from Eq (\ref{orbopt}) to update the Hamiltonian.
Finally, we continue the above steps to obtain $\lambda$ and $U$ self-consistently.
To efficiently perform the minimization, analytical energy derivatives with respect to either $\lambda$ or $X$ are required.
However, because of the mathematical complexity, we only compute derivatives with respect to $X$ numerically for the variational calculations.

\subsubsection{Low-rank AGP}

We also introduce the low-rank AGP wavefunction to compare it with the AGP and the low-rank APG.
The low-rank AGP has a formulation almost similar to that of low-rank APG[$\lambda, X$], but since all geminals in the AGP wavefunction are identical, there is no need to introduce the approximation that treats unitary matrices as identical.
Thus, it is of the simple form.
\begin{eqnarray}
\ket{\Psi_{\mathrm{r}x\mathrm{AGP}}} =  \left(\lambda^{(1)} \hat{a}^{\dag}_{1} \hat{a}^{\dag}_{\bar{1}} +\cdots  +\lambda^{(x)} \hat{a}^{\dag}_{x} \hat{a}^{\dag}_{\bar{x}}   \right)^{N/2} \ket{0}
\end{eqnarray}
Due to the Pauli's principle, note that the rank-$N/2$ AGP is equivalent to the rank-1 APG and the HF wavefunction.

\section{Demonstrative applications}

In this section, we present demonstrative applications of the APG, AGP, AGP-CI, low-rank APG, and low-rank AGP methods to one-dimensional Hubbard model with periodic boundary condition and small molecules.
First, we discuss the variational calculations of the APG, AGP, and AGP-CI wavefunctions to compare with the HF and the exact energies.
In addition, the Schur decompositions are carried out for these geminals to analyze eigenvalue spectra, from which we discuss how many eigenvalues mainly contribute to the ground state wavefunction.
Finally, we discuss the energies and wavefunctions of the low-rank APG and low-rank AGP, in comparisons to the APG and AGP, respectively.

\subsection{APG, AGP, and AGP-CI wavefunctions}

\subsubsection{Hubbard model}

\begin{figure}
\centering
{\resizebox*{6cm}{!}{\includegraphics{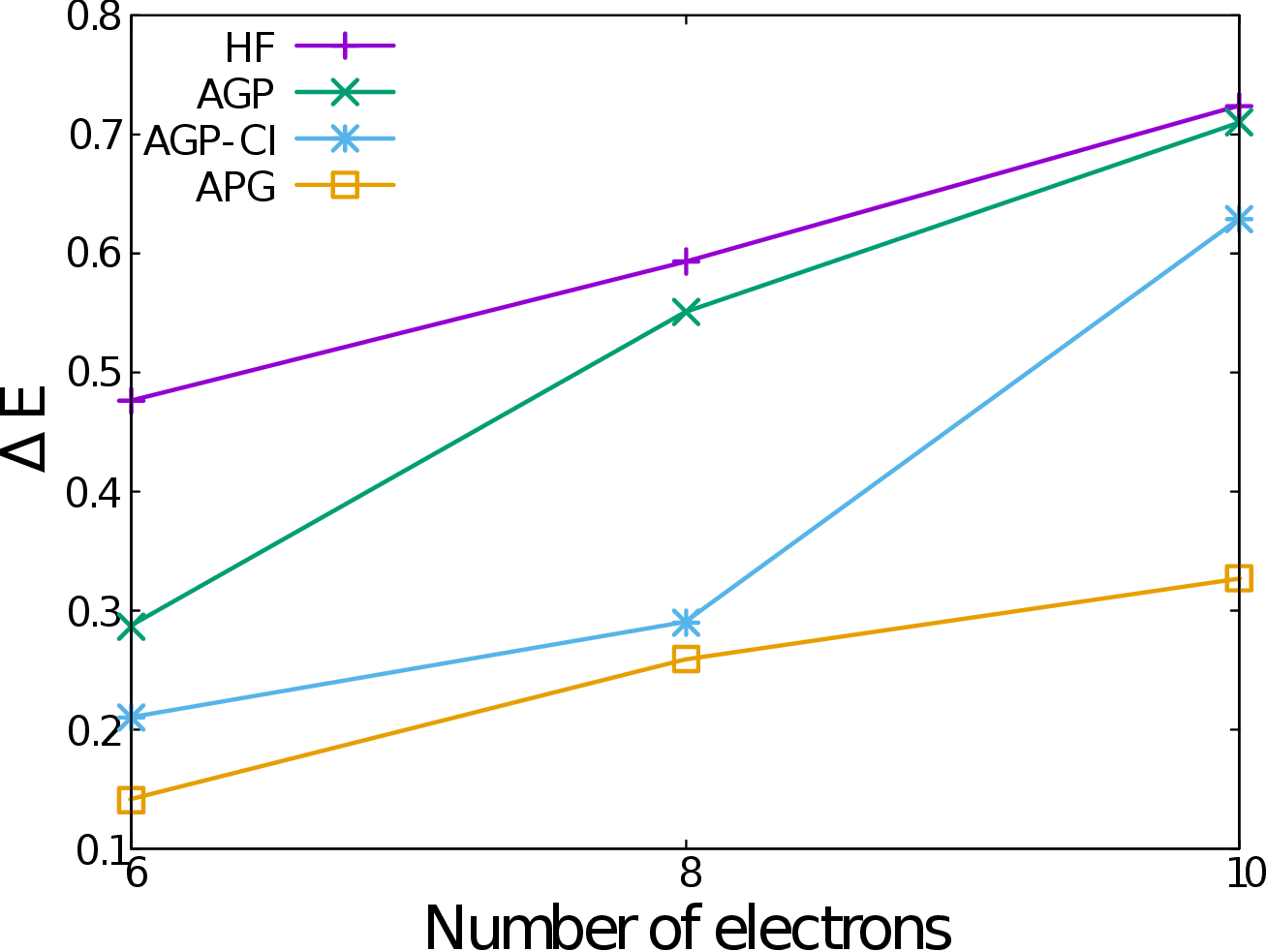}}}
\caption{Residual error of APG, AGP-CI, AGP and HF in the total energy of the Hubbard model ($U/t=10$ and half-filling) plotted against the number of electrons $N$.} 
\label{energy}
\end{figure}

Tables \ref{hubbard10} and \ref{hubbard1} showed the total energies of the periodic-boundary Hubbard models for $U/t=10$ and $U/t=1$, respectively.
The number of sites and electrons we considered were from 6 to 10.
In case of $U/t=10$ where electron correlation is strong, the APG method showed the smallest error from the exact solution, and the accuracy of the APG method is remarkably superior to that of other geminal methods.
Here, we took the AGP-CI to have the same number of variational parameters as in the APG wavefunction. Thus, it indicates that the APG method can represent the wavefunction more compactly.
In case of $U/t=1$, however, the APG and the AGP-CI results were almost the same.
In some cases, and this may be because the conjugate gradient optimization was not sufficient. Figure \ref{energy} summarizes the error in total energy for $U/t=10$ and half-filling.

\begin{table}
\tbl{Total energy of exact diagonalization, APG, AGP-CI, AGP and HF in the Hubbard model ($U/t=10$, $N$-site and $N$-electron system).}
{\begin{tabular}{lcccccccc} \toprule
(site, electron) & (6, 6) & (6, 8) & (8, 6) & (8, 8) & (8, 10) & (10, 6) & (10, 8) & (10, 10) \\ \midrule
 Exact & -1.66436 & 15.92144 & -5.17812 & -2.17669 & 14.82188 & -7.23006 & -5.66976 & -2.70369 \\
 APG & -1.52224 & 15.92325 & -4.44375 & -1.91733 & 15.40734 & -6.55893 & -4.82093 & -2.37648 \\
 AGP-CI & -1.45352 & 15.92359 & -4.29550 & -1.88584 & 15.62349 & -6.53923 & -4.64035 & -2.07456 \\
 AGP & -1.23490 & 16.36844 & -3.92302 & -1.62613 & 16.24880 & -6.19103 & -4.16048 & -1.99416 \\ 
 HF & -1.18824 & 16.62690 & -3.70321 & -1.58400 & 16.29679 & -6.07635 & -4.04319 & -1.98001 \\  \bottomrule
\end{tabular}}
\label{hubbard10}
\end{table}

\begin{table}
\tbl{Total energy of exact diagonalization, APG, AGP-CI, AGP and HF in the Hubbard model ($U/t=1$, $N$-site and $N$-electron system). }
{\begin{tabular}{lcccccccc} \toprule
(site, electron) & (6, 6) & (6, 8) & (8, 6) & (8, 8) & (8, 10) & (10, 6) & (10, 8) & (10, 10) \\ \midrule
 Exact & -6.60116 & -3.54928 & -8.64163 & -7.95233 & -6.64163 & -9.68212 & -10.33943 & -10.61441 \\
 APG & -6.59603 & -3.54921 & -8.62678 & -7.91695 & -6.63507 & -9.66203 & -10.31342 & -10.58208 \\
 AGP-CI & -6.59296 & -3.54925 & -8.62035 & -7.91457 & -6.63524 & -9.66164 & -10.31861 &  -10.58291\\
 AGP & -6.54373 & -3.51756 & -8.57615 & -7.84228 & -6.55994 & -9.61574 & -10.23356 & -10.48433 \\
 HF & -6.50000 & -3.50000 & -8.53185 & -7.82102 & -6.53185 & -9.57214 & -10.20820 & -10.44427 \\  \bottomrule
\end{tabular}}
\label{hubbard1}
\end{table}

\begin{figure}
\centering
{%
\resizebox*{4.5cm}{!}{\includegraphics{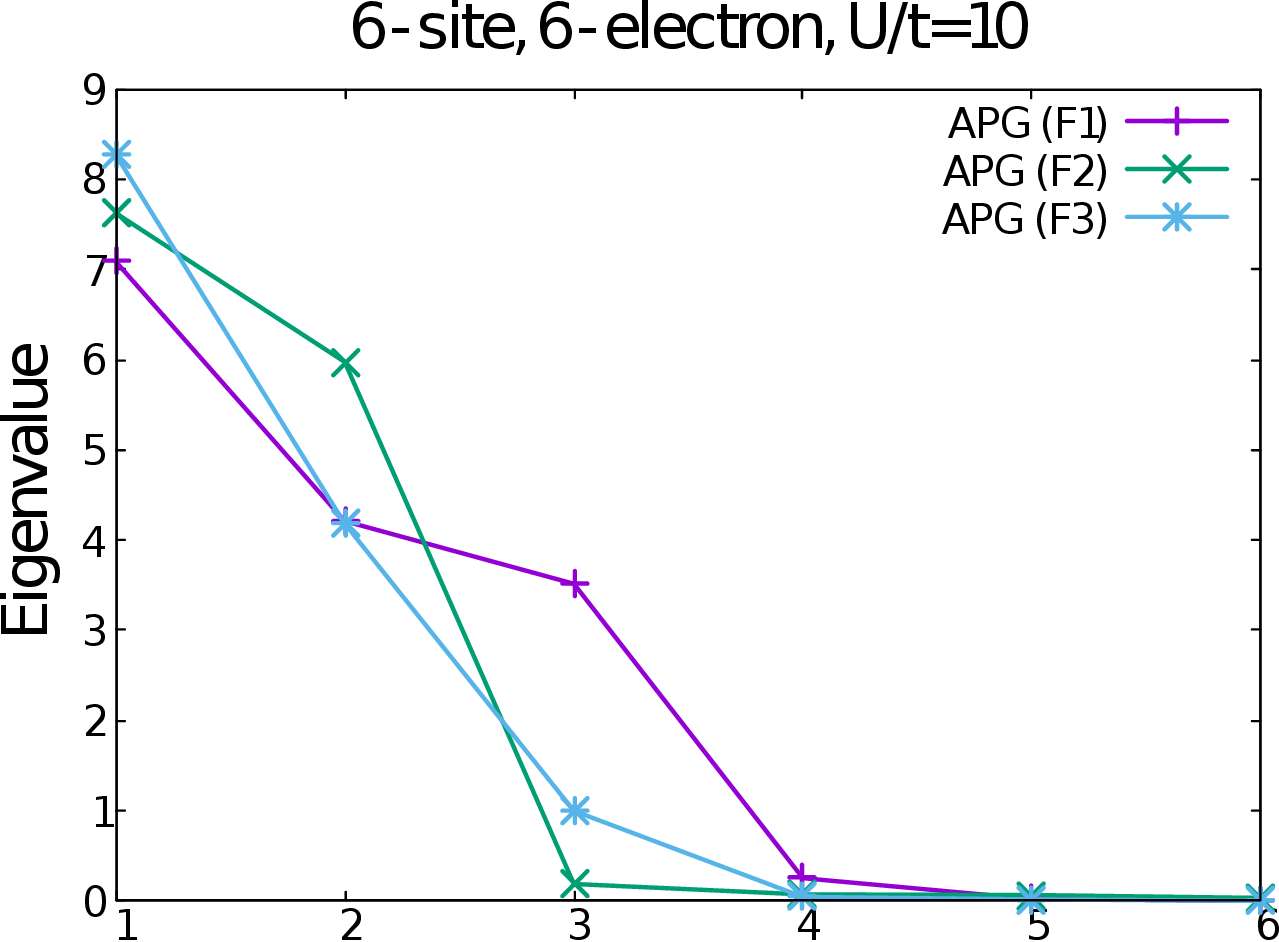}}}\hspace{5pt}
{%
\resizebox*{4.5cm}{!}{\includegraphics{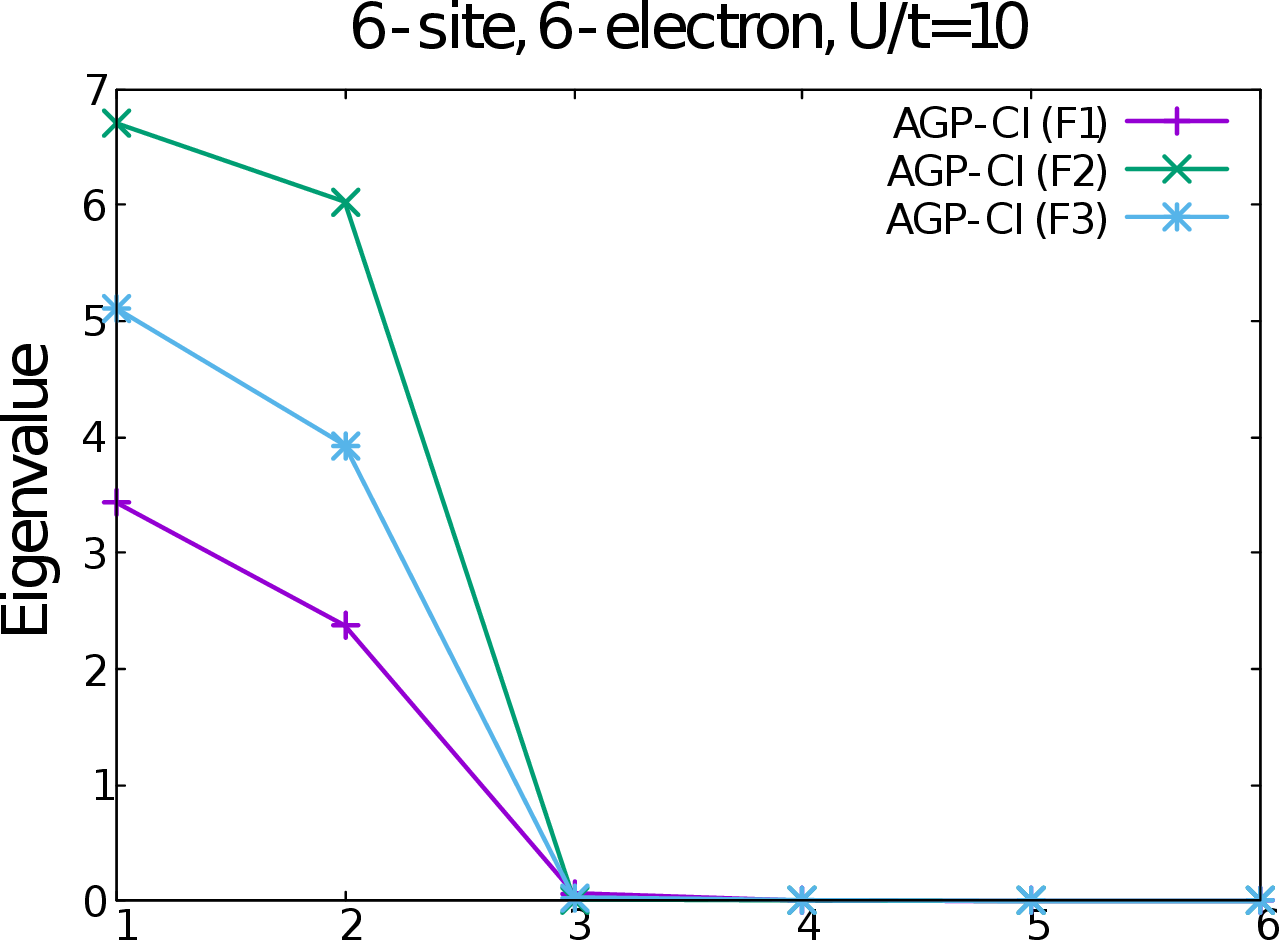}}}\hspace{5pt} 
{%
\resizebox*{4.5cm}{!}{\includegraphics{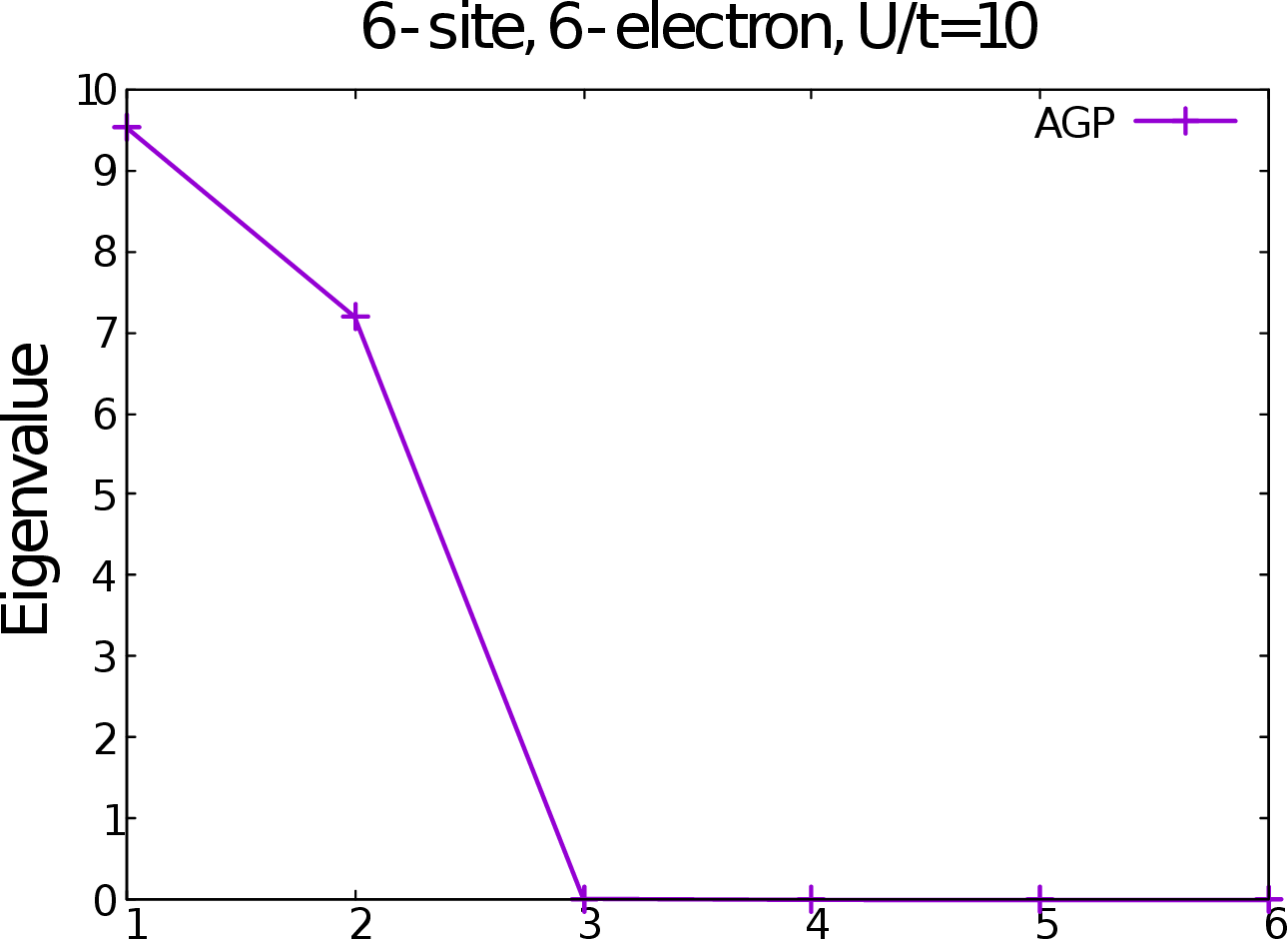}}} \vspace{10pt} \\
{%
\resizebox*{4.5cm}{!}{\includegraphics{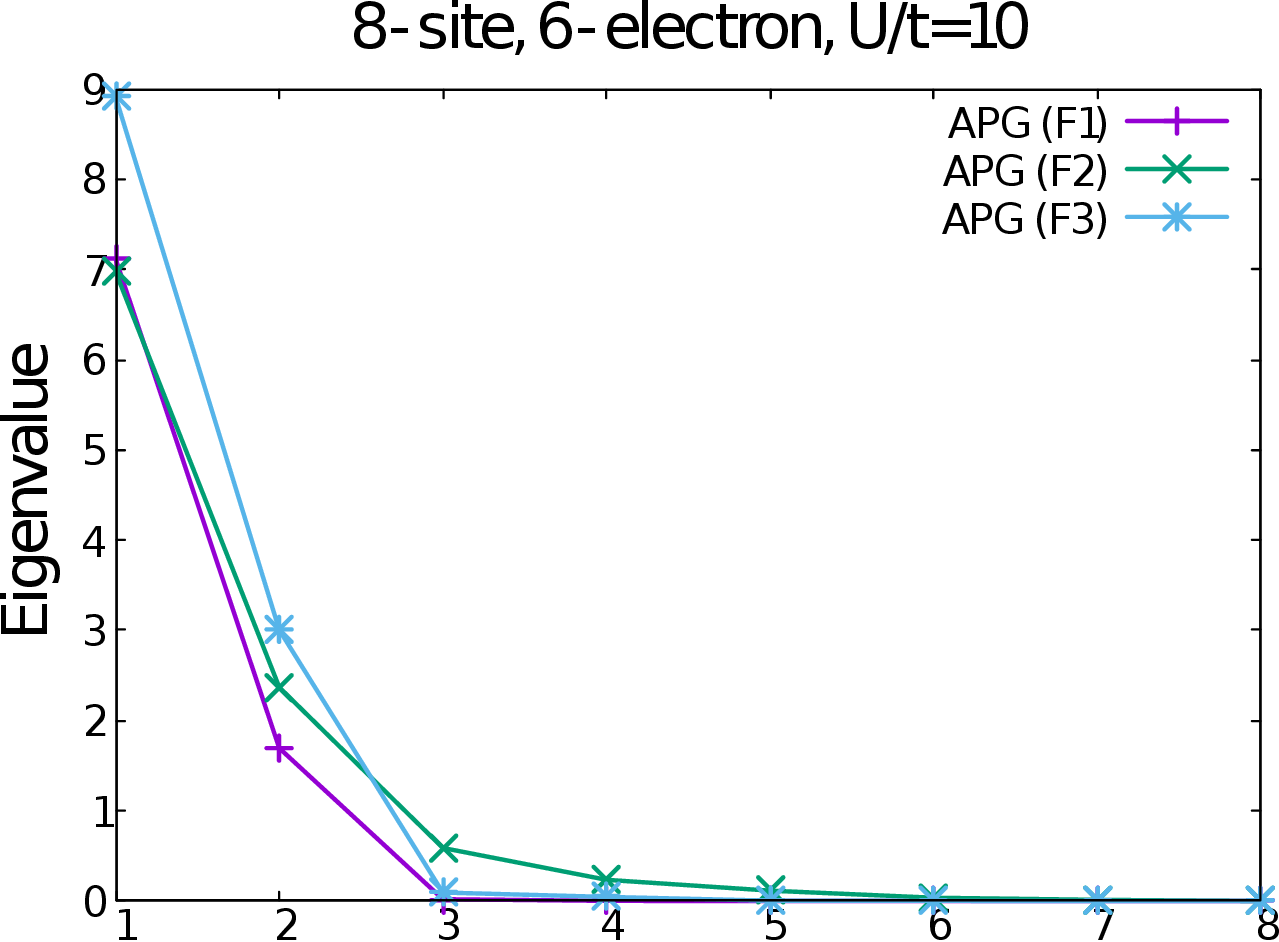}}}\hspace{5pt}
{%
\resizebox*{4.5cm}{!}{\includegraphics{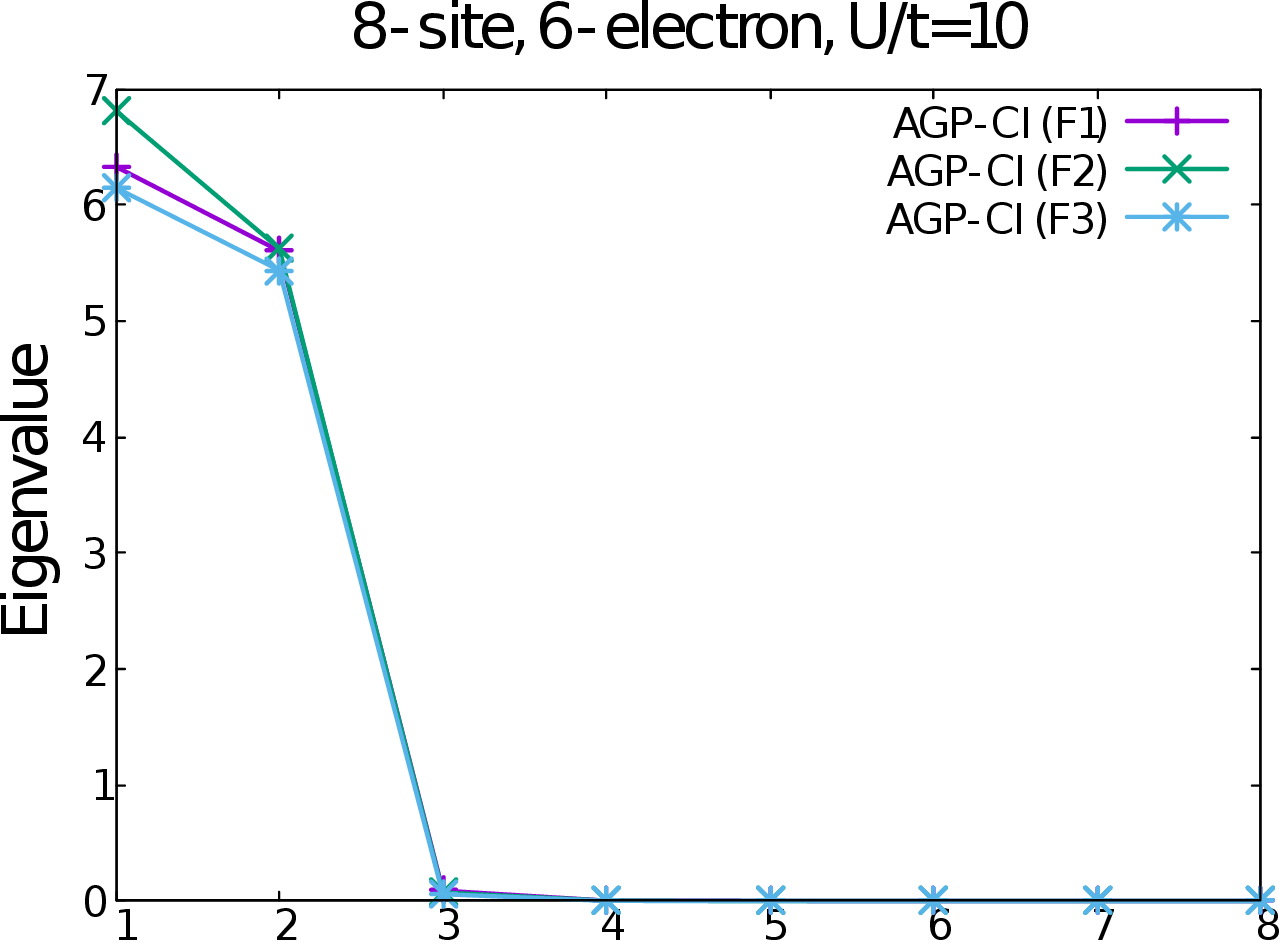}}}\hspace{5pt}
{%
\resizebox*{4.5cm}{!}{\includegraphics{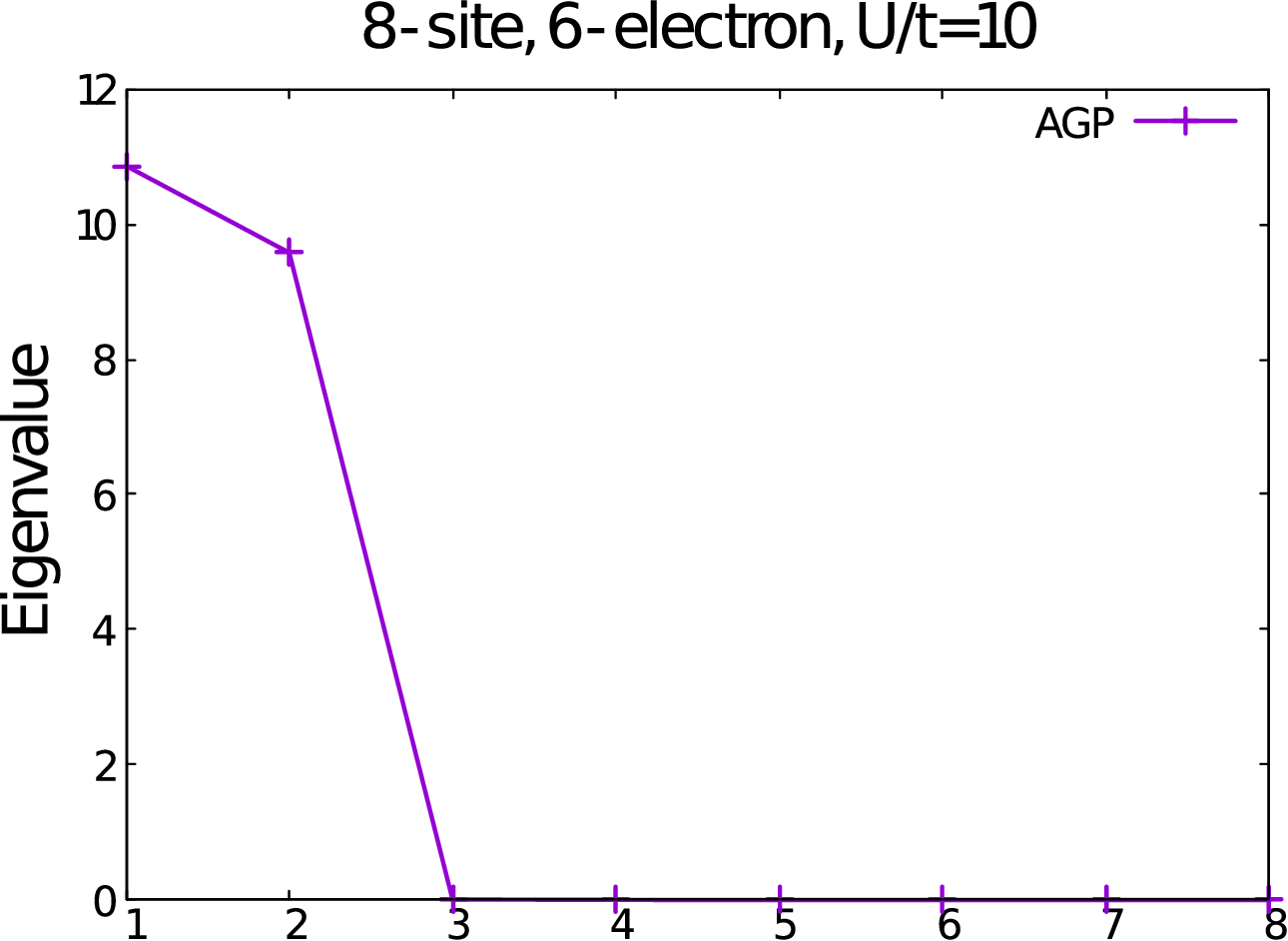}}} \vspace{10pt} \\
{%
\resizebox*{4.5cm}{!}{\includegraphics{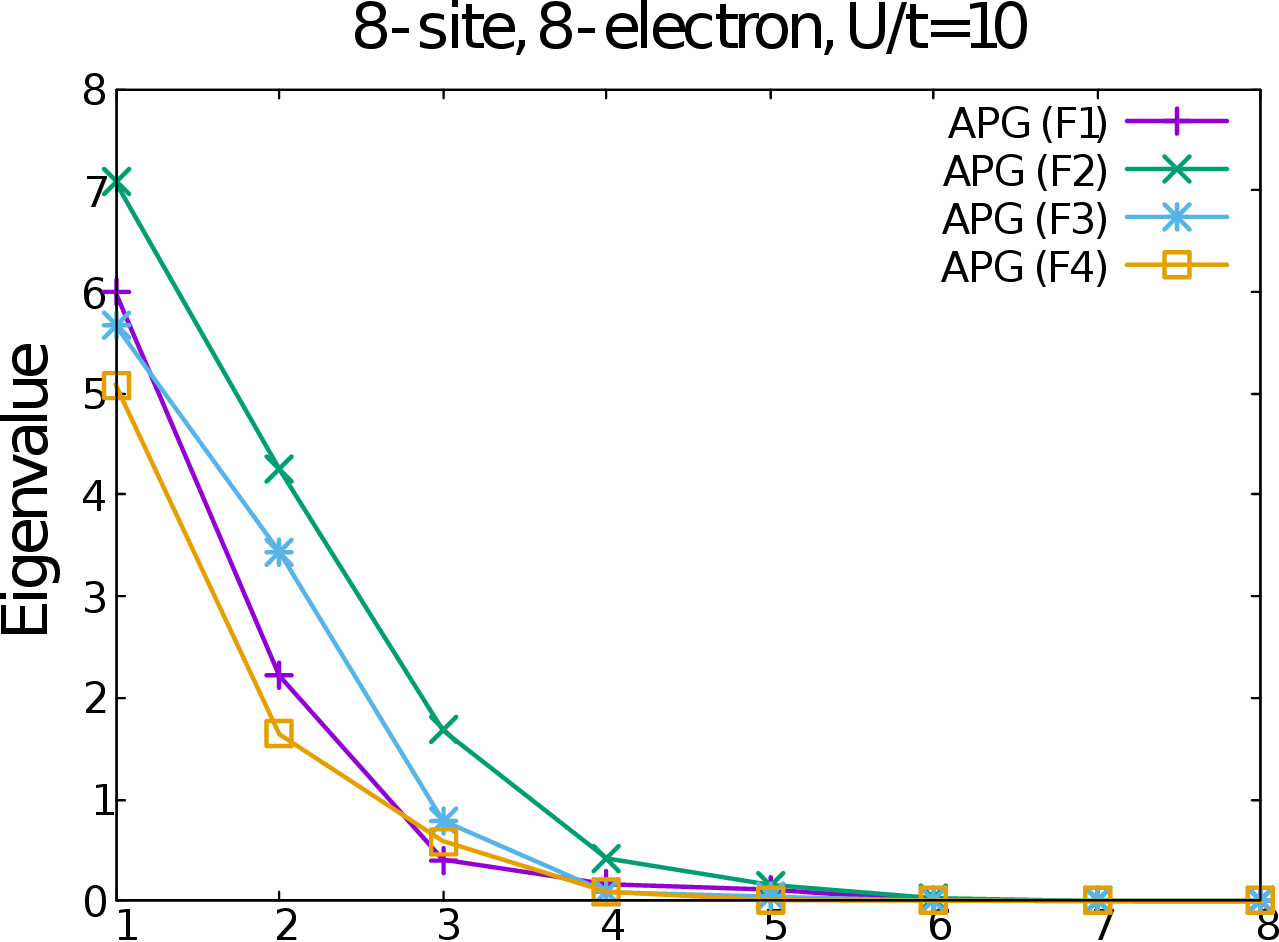}}}\hspace{5pt}
{%
\resizebox*{4.5cm}{!}{\includegraphics{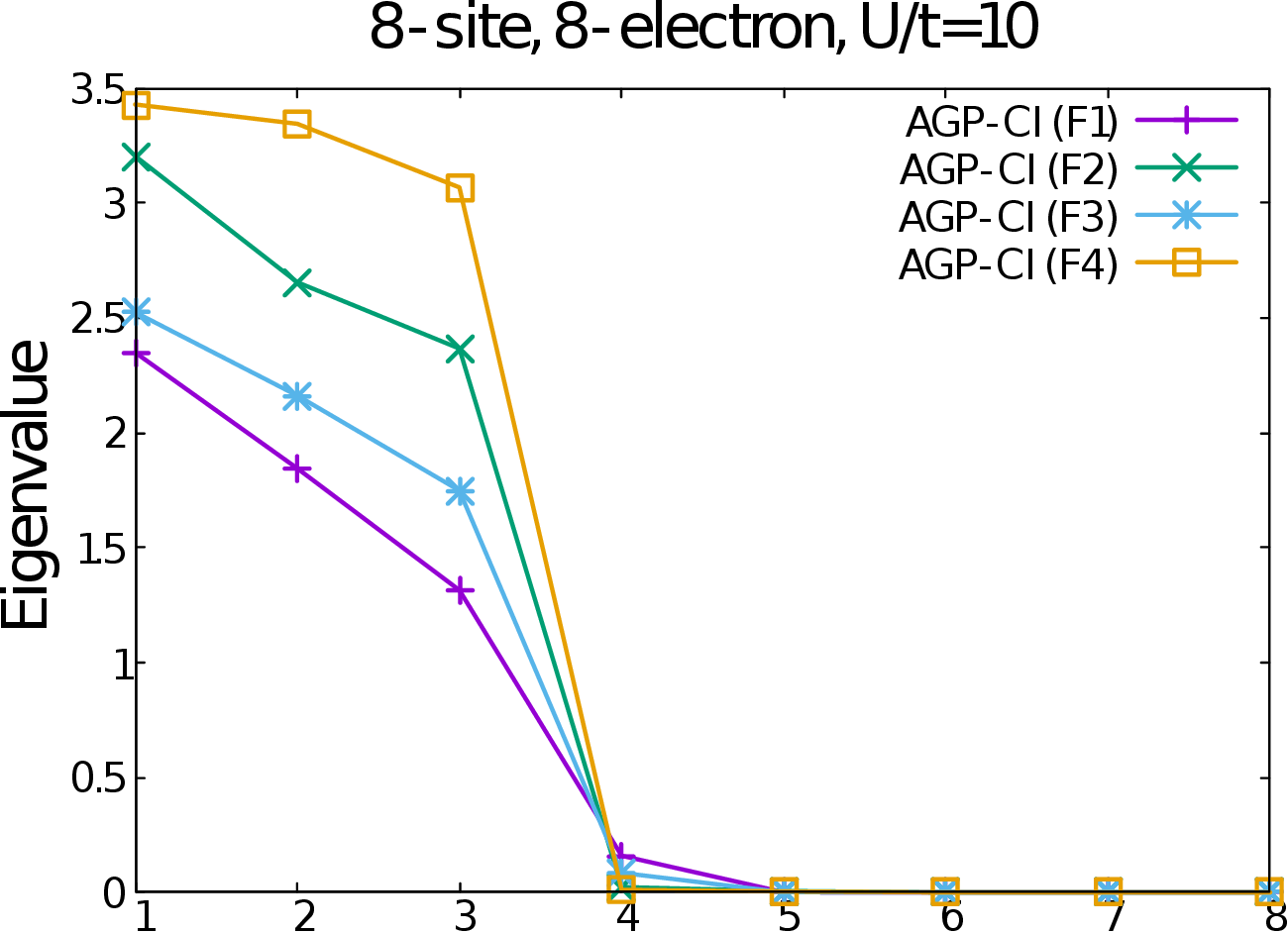}}}\hspace{5pt}
{%
\resizebox*{4.5cm}{!}{\includegraphics{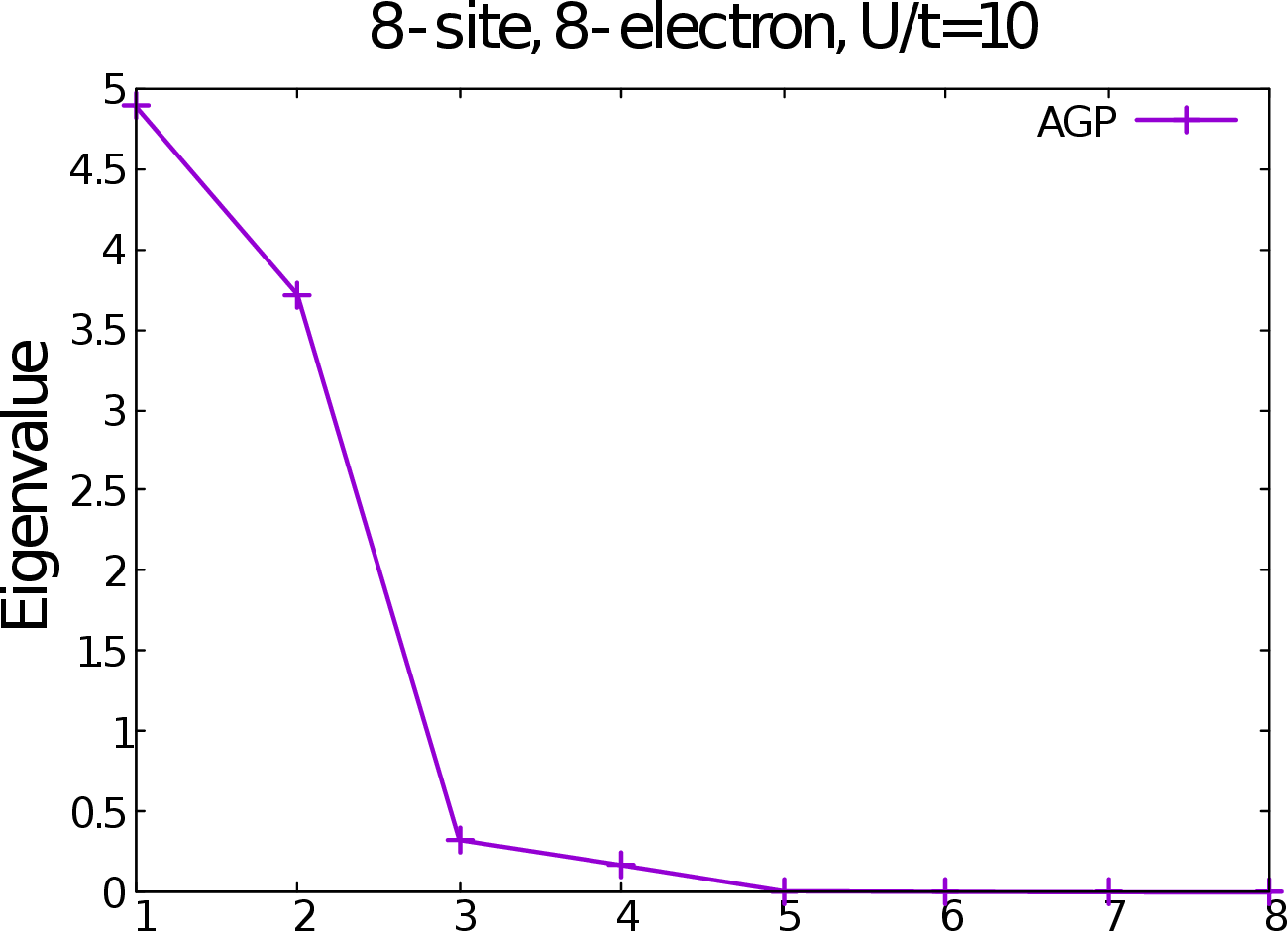}}}  \\
\subfloat[APG]{%
\resizebox*{4.5cm}{!}{\includegraphics{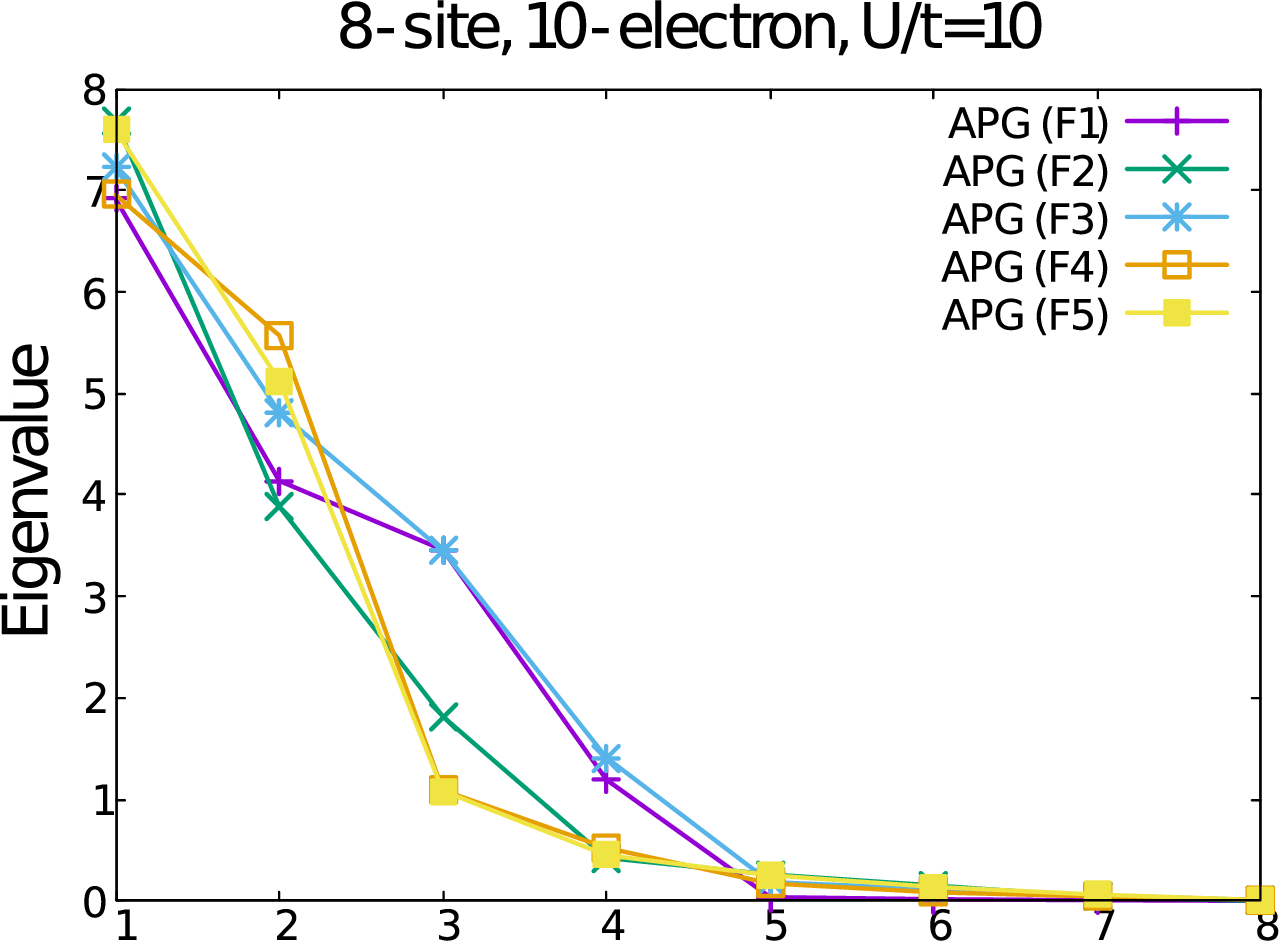}}}\hspace{5pt}
\subfloat[AGP-CI]{%
\resizebox*{4.5cm}{!}{\includegraphics{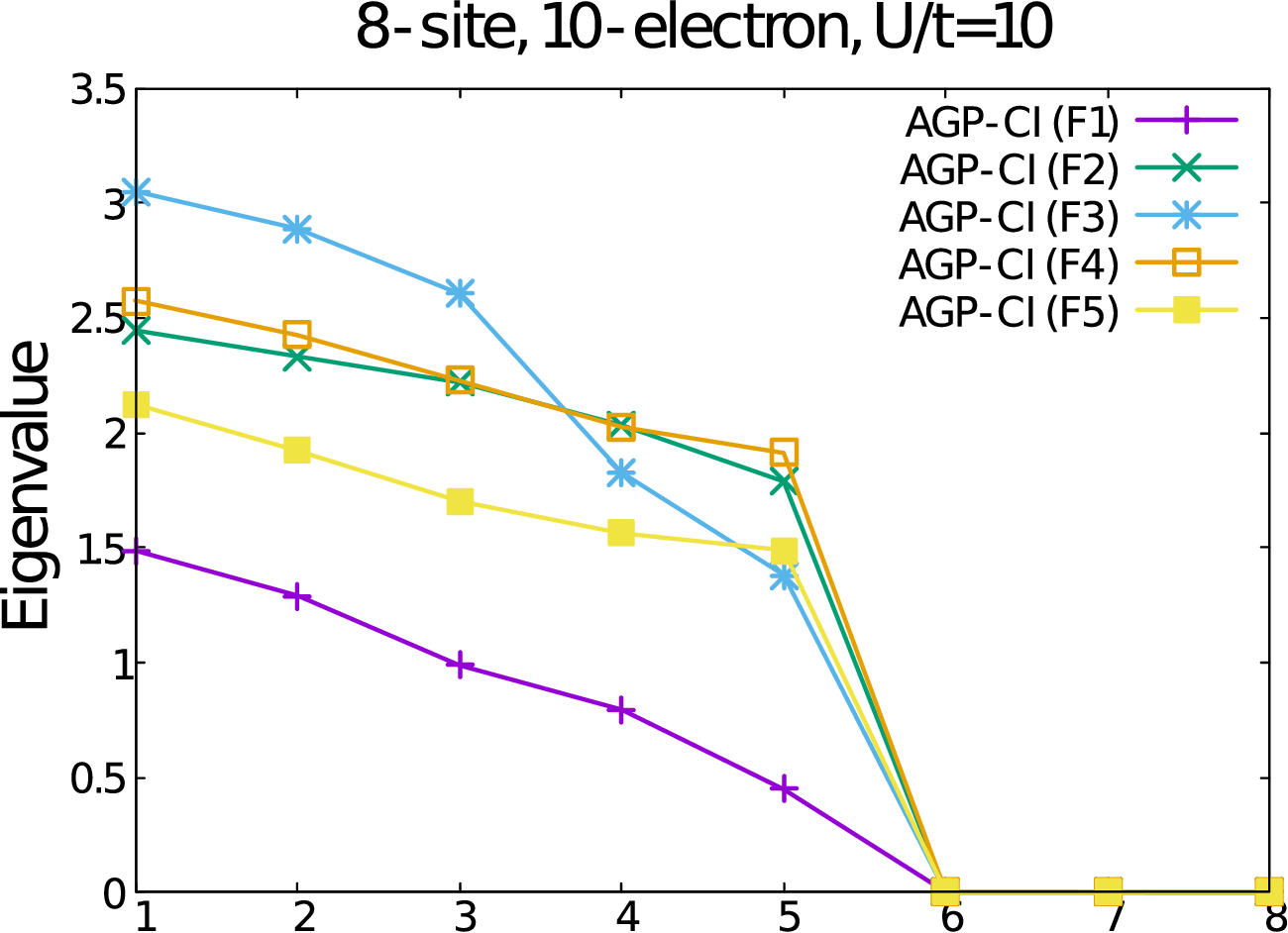}}}\hspace{5pt}
\subfloat[AGP]{%
\resizebox*{4.5cm}{!}{\includegraphics{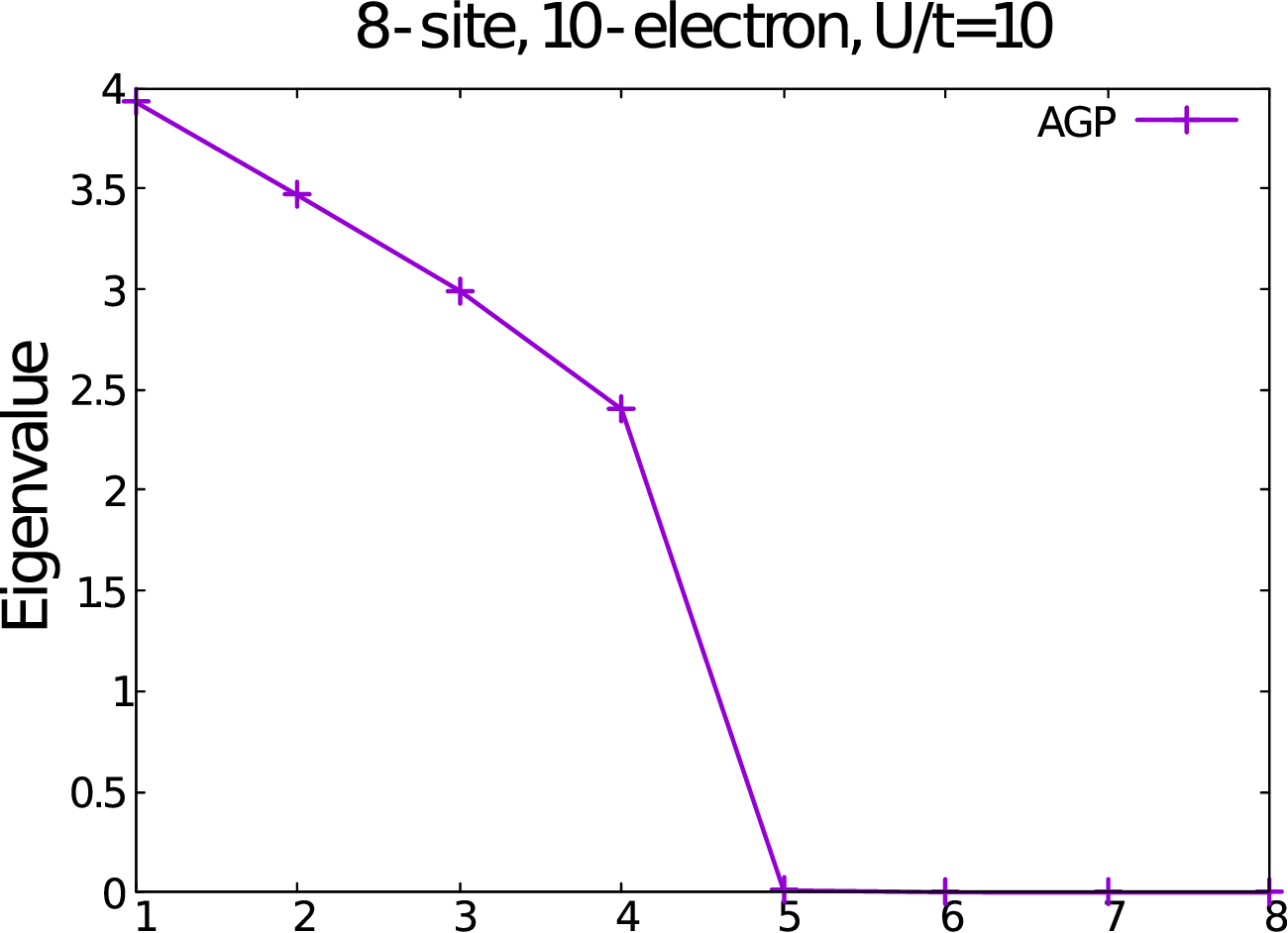}}}
\caption{The absolute eigenvalues of geminal matrices in (a) APG, (b) AGP-CI and (c) AGP in the Hubbard model ($U/t=10$).} 
\label{hubbard_eigen}
\end{figure}

Furthermore, we analyzed the geminal matrices in the geminal wavefunctions in order to explore the nature of the electron correlation.
Geminal skew-symmetric matrices can be band-diagonalized to get the eigenvalues as in Eq (\ref{usu}) and (\ref{diag}).
Figure \ref{hubbard_eigen} showed the absolute eigenvalues of each of the geminal matrices after optimization, arranged in descending order.
It can be seen that there are many eigenvalues that are almost zero.
Comparing the upper two rows of Figure \ref{hubbard_eigen}, the systems containing 6 electrons, number of non-zero eigenvalues is always 3 even if number of sites changes.
Comparing the bottom three rows, the systems containing 6, 8 and 10 electrons, numbers of non-zero eigenvalues are 3, 4 and 5, respectively.
Consequently, it is concluded that number of non-zero eigenvalues depends only on number of electrons, which is about $N/2$ for APG and AGP-CI, and $N/2-1$ for AGP.

\subsubsection{Small molecules}

Table \ref{molecule} showed the total energies of H$_6$ chain, Li$_{2}$, and H$_2$O computed by APG, AGP-CI, and AGP, compared with the HF and the exact energies using STO-6G basis sets.
In H$_6$ chain, the APG energy was computed to be very close to the exact one.
The AGP-CI also showed good performance, but it gave slightly higher energy than either the APG or the exact solution.
While in Li$_2$ and H$_2$O molecules, both the APG and the AGP-CI energies were computed to be very close to the exact one.
These results indicate that the APG can sufficiently incorporate strong electron correlations, but the AGP-CI cannot fully incorporate them.

\begin{figure}
\centering
\subfloat[APG]{%
\resizebox*{4.5cm}{!}{\includegraphics{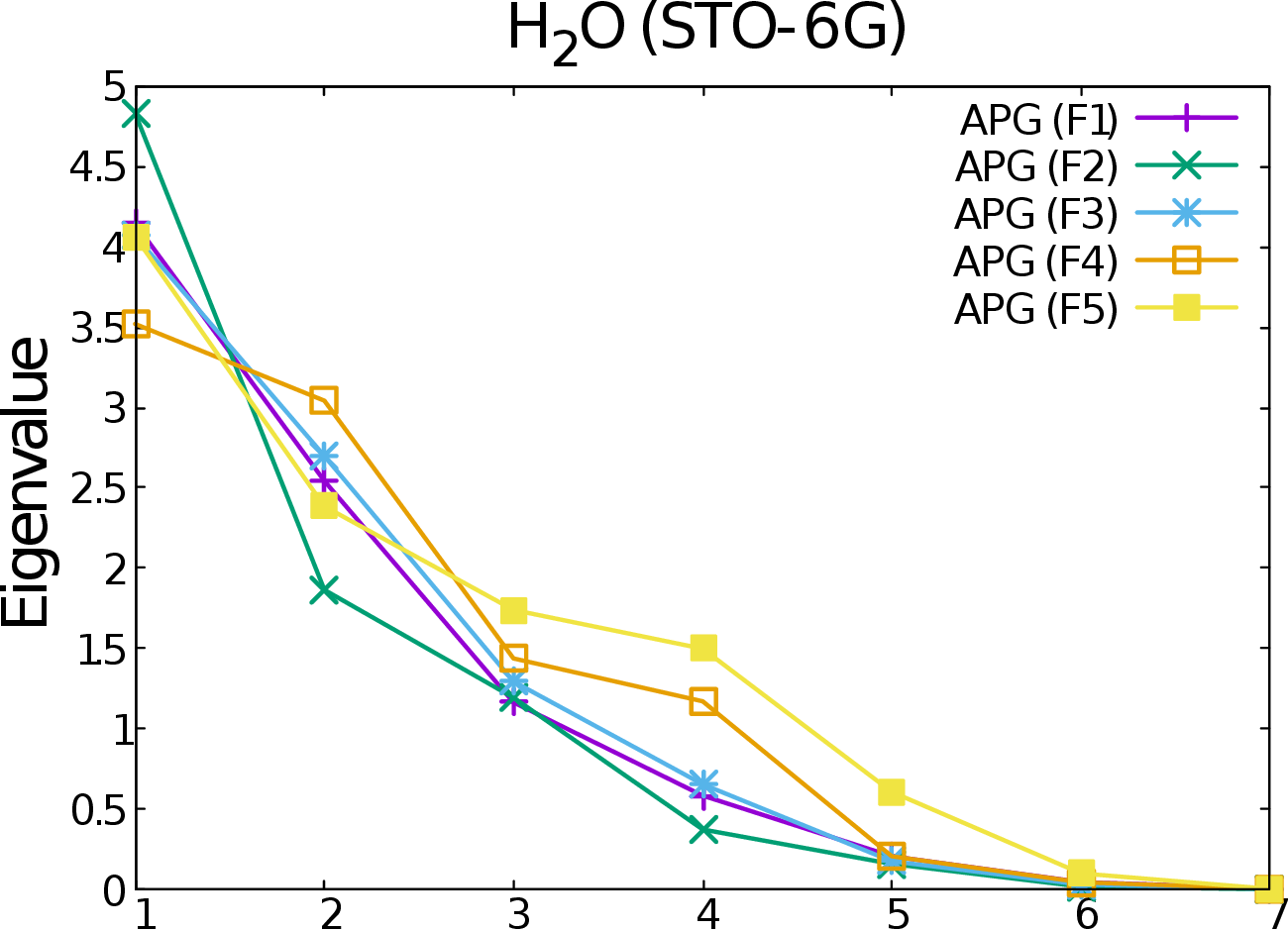}}}\hspace{5pt}
\subfloat[AGP-CI]{%
\resizebox*{4.5cm}{!}{\includegraphics{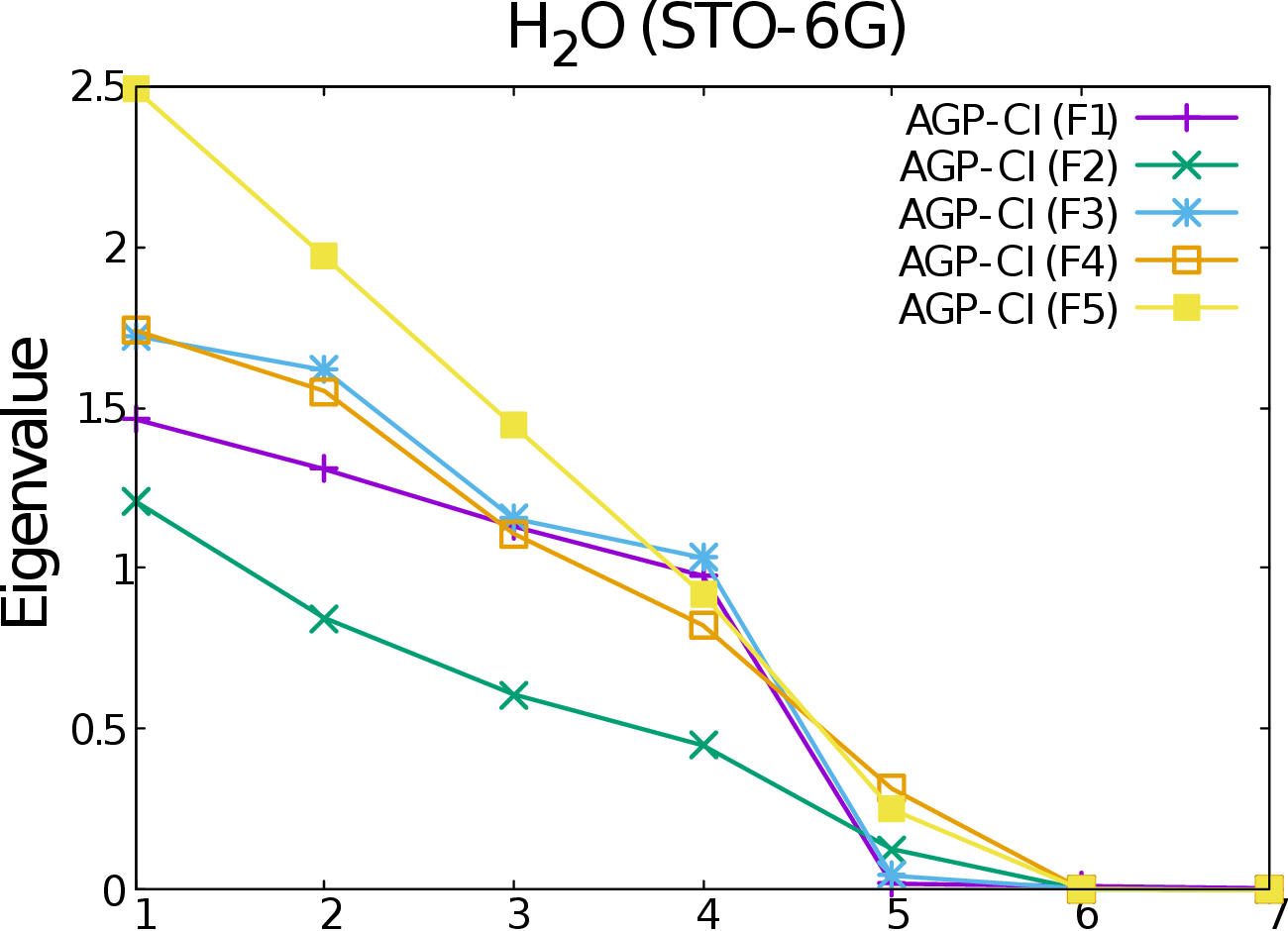}}}\hspace{5pt}
\subfloat[AGP]{%
\resizebox*{4.5cm}{!}{\includegraphics{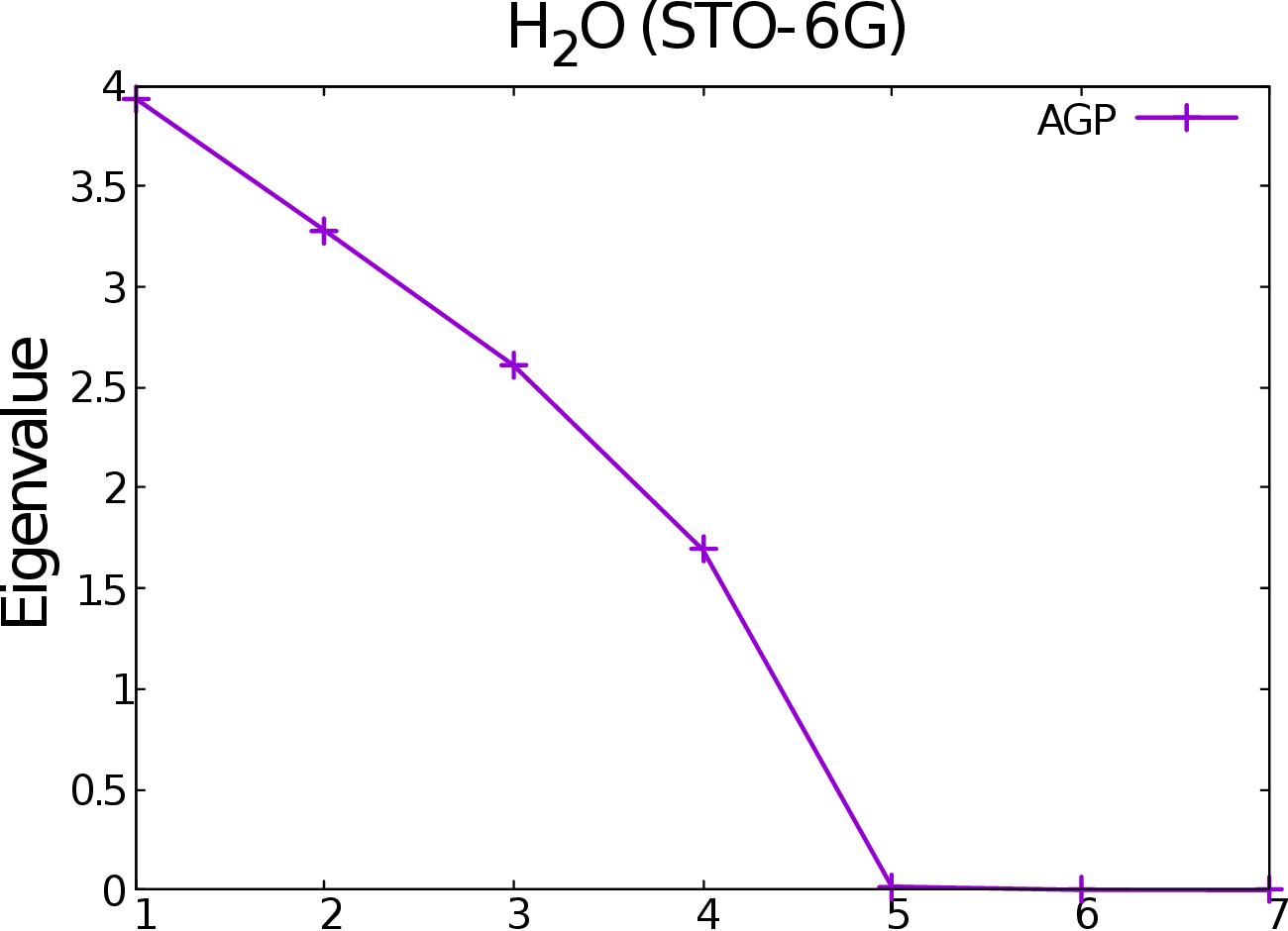}}}
\caption{The absolute eigenvalues of geminal matrices in (a) APG, (b) AGP-CI and (c) AGP in H$_{2}$O (STO-6G).} 
\label{h2o_eigen}
\end{figure}

Figure \ref{h2o_eigen} showed the absolute eigenvalues of each geminal matrix in H$_2$O calculations, arranged in descending order.
Similarly to the Hubbard model, number of non-zero eigenvalues are up to $N/2=5$ for APG and AGP-CI, and $N/2-1=4$ for AGP in H$_2$O molecule ($N=10$).

\begin{table}
\tbl{Total energy of exact diagonalization, APG, AGP-CI, AGP and HF in hydrogen chain (H$_6$), Li$_{2}$ and H$_2$O (STO-6G).}
{\begin{tabular}{lccccc} \toprule
 Total energy (Hartree) &  & H$_6$ & Li$_{2}$ & H$_2$O   \\ \midrule
 Exact &   & -3.173108  & -14.837627  &-75.728706  \\
 APG &   &  -3.170922 &  -14.837181 &-75.728417  \\
 AGP-CI & &  -3.167128 &  -14.837130 &  -75.728416  \\
 AGP &   &  -3.132869 &  -14.836786 & -75.693839\\
 HF &  & -3.110464  &  -14.811204 & -75.678686  \\  \bottomrule
\end{tabular}}
\label{molecule}
\end{table}

\subsection{Low-rank APG} \label{low-rank_x}

Table \ref{low-rank_x1} and \ref{low-rank_x2} showed the total energies obtained from the low-rank APG's with the Hubbard model ($U/t=10$ and $U/t=1$, 6-site and 6-electron system).
To show how well the low-rank APG's can reproduce APG, we also included coverages of the correlation energy of APG.
Since various forms of wavefunctions can be taken after rank-2 APG, the following wavefunctions were used in the calculations in this paper.
\begin{eqnarray}
\ket{\Psi_{\mathrm{r1APG}}} &=& \lambda[1]^{(1)} \hat{a}^{\dag}_{1} \hat{a}^{\dag}_{\bar{1}} \lambda[2]^{(2)} \hat{a}^{\dag}_{2} \hat{a}^{\dag}_{\bar{2}} \lambda[3]^{(3)} \hat{a}^{\dag}_{3} \hat{a}^{\dag}_{\bar{3}} \ket{0} \\
\ket{\Psi_{\mathrm{r2APG (a)}}} &=& \left(\lambda[1]^{(1)} \hat{a}^{\dag}_{1} \hat{a}^{\dag}_{\bar{1}} +\lambda[1]^{(2)} \hat{a}^{\dag}_{2} \hat{a}^{\dag}_{\bar{2}}  \right)\left(\lambda[2]^{(3)} \hat{a}^{\dag}_{3} \hat{a}^{\dag}_{\bar{3}} +\lambda[2]^{(4)} \hat{a}^{\dag}_{4} \hat{a}^{\dag}_{\bar{4}}  \right)\left(\lambda[3]^{(5)} \hat{a}^{\dag}_{5} \hat{a}^{\dag}_{\bar{5}} +\lambda[3]^{(6)} \hat{a}^{\dag}_{6} \hat{a}^{\dag}_{\bar{6}}  \right) \ket{0} \nonumber \\
&&\\
\ket{\Psi_{\mathrm{r2APG (b)}}} &=& \left(\lambda[1]^{(1)} \hat{a}^{\dag}_{1} \hat{a}^{\dag}_{\bar{1}} +\lambda[1]^{(2)} \hat{a}^{\dag}_{2} \hat{a}^{\dag}_{\bar{2}}  \right)\left(\lambda[2]^{(3)} \hat{a}^{\dag}_{3} \hat{a}^{\dag}_{\bar{3}} +\lambda[2]^{(4)} \hat{a}^{\dag}_{4} \hat{a}^{\dag}_{\bar{4}}  \right)\left(\lambda[3]^{(5)} \hat{a}^{\dag}_{5} \hat{a}^{\dag}_{\bar{5}} +\lambda[3]^{(1)} \hat{a}^{\dag}_{1} \hat{a}^{\dag}_{\bar{1}}  \right) \ket{0} \nonumber \\
&&\\
\ket{\Psi_{\mathrm{r2APG (c)}}} &=& \left(\lambda[1]^{(1)} \hat{a}^{\dag}_{1} \hat{a}^{\dag}_{\bar{1}} +\lambda[1]^{(2)} \hat{a}^{\dag}_{2} \hat{a}^{\dag}_{\bar{2}}  \right)\left(\lambda[2]^{(3)} \hat{a}^{\dag}_{3} \hat{a}^{\dag}_{\bar{3}} +\lambda[2]^{(4)} \hat{a}^{\dag}_{4} \hat{a}^{\dag}_{\bar{4}}  \right)\left(\lambda[3]^{(1)} \hat{a}^{\dag}_{1} \hat{a}^{\dag}_{\bar{1}} +\lambda[3]^{(3)} \hat{a}^{\dag}_{3} \hat{a}^{\dag}_{\bar{3}}  \right) \ket{0} \nonumber \\
&&\\
\ket{\Psi_{\mathrm{r3APG}}} &=& \left(\lambda[1]^{(1)} \hat{a}^{\dag}_{1} \hat{a}^{\dag}_{\bar{1}} +\lambda[1]^{(4)} \hat{a}^{\dag}_{4} \hat{a}^{\dag}_{\bar{4}} +\lambda[1]^{(2)} \hat{a}^{\dag}_{2} \hat{a}^{\dag}_{\bar{2}}  \right)\left(\lambda[2]^{(2)} \hat{a}^{\dag}_{2} \hat{a}^{\dag}_{\bar{2}} +\lambda[2]^{(5)} \hat{a}^{\dag}_{5} \hat{a}^{\dag}_{\bar{5}} +\lambda[2]^{(3)} \hat{a}^{\dag}_{3} \hat{a}^{\dag}_{\bar{3}}  \right) \nonumber \\
&&\left(\lambda[3]^{(3)} \hat{a}^{\dag}_{3} \hat{a}^{\dag}_{\bar{3}} +\lambda[3]^{(6)} \hat{a}^{\dag}_{6} \hat{a}^{\dag}_{\bar{6}} +\lambda[3]^{(1)} \hat{a}^{\dag}_{1} \hat{a}^{\dag}_{\bar{1}}  \right) \ket{0} \\
\ket{\Psi_{\mathrm{r4APG}}} &=& \left(\lambda[1]^{(1)} \hat{a}^{\dag}_{1} \hat{a}^{\dag}_{\bar{1}} +\lambda[1]^{(2)} \hat{a}^{\dag}_{2} \hat{a}^{\dag}_{\bar{2}} +\lambda[1]^{(3)} \hat{a}^{\dag}_{3} \hat{a}^{\dag}_{\bar{3}} +\lambda[1]^{(4)} \hat{a}^{\dag}_{4} \hat{a}^{\dag}_{\bar{4}}  \right)\nonumber \\
&&\left(\lambda[2]^{(1)} \hat{a}^{\dag}_{1} \hat{a}^{\dag}_{\bar{1}} +\lambda[2]^{(2)} \hat{a}^{\dag}_{2} \hat{a}^{\dag}_{\bar{2}} +\lambda[2]^{(5)} \hat{a}^{\dag}_{5} \hat{a}^{\dag}_{\bar{5}}+\lambda[2]^{(6)} \hat{a}^{\dag}_{6} \hat{a}^{\dag}_{\bar{6}}   \right) \nonumber \\
&&\left(\lambda[3]^{(3)} \hat{a}^{\dag}_{3} \hat{a}^{\dag}_{\bar{3}} +\lambda[3]^{(4)} \hat{a}^{\dag}_{4} \hat{a}^{\dag}_{\bar{4}} +\lambda[3]^{(5)} \hat{a}^{\dag}_{5} \hat{a}^{\dag}_{\bar{5}} +\lambda[3]^{(6)} \hat{a}^{\dag}_{6} \hat{a}^{\dag}_{\bar{6}}   \right) \ket{0} \\
\ket{\Psi_{\mathrm{r5APG}}} &=& \left(\lambda[1]^{(1)} \hat{a}^{\dag}_{1} \hat{a}^{\dag}_{\bar{1}} +\lambda[1]^{(2)} \hat{a}^{\dag}_{2} \hat{a}^{\dag}_{\bar{2}} +\lambda[1]^{(3)} \hat{a}^{\dag}_{3} \hat{a}^{\dag}_{\bar{3}} +\lambda[1]^{(4)} \hat{a}^{\dag}_{4} \hat{a}^{\dag}_{\bar{4}}+\lambda[1]^{(5)} \hat{a}^{\dag}_{5} \hat{a}^{\dag}_{\bar{5}}    \right)\nonumber \\
&&\left(\lambda[2]^{(1)} \hat{a}^{\dag}_{1} \hat{a}^{\dag}_{\bar{1}} +\lambda[2]^{(2)} \hat{a}^{\dag}_{2} \hat{a}^{\dag}_{\bar{2}} +\lambda[2]^{(3)} \hat{a}^{\dag}_{3} \hat{a}^{\dag}_{\bar{3}}+\lambda[2]^{(4)} \hat{a}^{\dag}_{4} \hat{a}^{\dag}_{\bar{4}}  +\lambda[2]^{(6)} \hat{a}^{\dag}_{6} \hat{a}^{\dag}_{\bar{6}}  \right) \nonumber \\
&&\left(\lambda[3]^{(1)} \hat{a}^{\dag}_{1} \hat{a}^{\dag}_{\bar{1}} +\lambda[3]^{(2)} \hat{a}^{\dag}_{2} \hat{a}^{\dag}_{\bar{2}} +\lambda[3]^{(3)} \hat{a}^{\dag}_{3} \hat{a}^{\dag}_{\bar{3}} +\lambda[3]^{(5)} \hat{a}^{\dag}_{5} \hat{a}^{\dag}_{\bar{5}}  +\lambda[3]^{(6)} \hat{a}^{\dag}_{6} \hat{a}^{\dag}_{\bar{6}}   \right) \ket{0} \\
\ket{\Psi_{\mathrm{r6APG}}} &=& \left(\lambda[1]^{(1)} \hat{a}^{\dag}_{1} \hat{a}^{\dag}_{\bar{1}} +\cdots+\lambda[1]^{(6)} \hat{a}^{\dag}_{6} \hat{a}^{\dag}_{\bar{6}}    \right)\left(\lambda[2]^{(1)} \hat{a}^{\dag}_{1} \hat{a}^{\dag}_{\bar{1}}+\cdots  +\lambda[2]^{(6)} \hat{a}^{\dag}_{6} \hat{a}^{\dag}_{\bar{6}}  \right) \nonumber \\
&&\left(\lambda[3]^{(1)} \hat{a}^{\dag}_{1} \hat{a}^{\dag}_{\bar{1}} +\cdots  +\lambda[3]^{(6)} \hat{a}^{\dag}_{6} \hat{a}^{\dag}_{\bar{6}}   \right) \ket{0}
\end{eqnarray}

\begin{table}
\tbl{Total energy and the percentage of correlation energy ($E_{c}=E-E^{\mathrm{HF}}$) of exact diagonalization, APG, low-rank APG[$\lambda, X$] and HF in the Hubbard model ($U/t=10$, 6-site and 6-electron system).}
{\begin{tabular}{lcc} \toprule
  $U/t=10$      &Total energy ($/t$)& $(E_c/E_{c}^{\mathrm{APG}})*100$ ($\%$)  \\ \midrule
 Exact & -1.664363 &  - \\
 APG   & -1.522237 &  100 \\
 Rank-6 APG& -1.406358 &  65.3049 \\
 Rank-5 APG& -1.406312 &  65.2911\\
 Rank-4 APG& -1.406047 &  65.2118  \\
 Rank-3 APG& -1.406046 &  65.2116  \\
 Rank-2 (a) APG& -1.406045 &   65.2113 \\
 Rank-2 (b) APG& -1.337574 &   44.7108 \\
 Rank-2 (c) APG& -1.263858 &   22.6395 \\
 Rank-1 APG& -1.188243 &    0\\
 HF   & -1.188243 &   0 \\ \bottomrule
\end{tabular}}
\label{low-rank_x1}
\end{table}

\begin{table}
\tbl{Total energy and the percentage of correlation energy ($E_{c}=E-E^{\mathrm{HF}}$) of exact diagonalization, APG, low-rank APG[$\lambda, X$] and HF in the Hubbard model ($U/t=1$, 6-site and 6-electron system).}
{\begin{tabular}{lcc} \toprule
  $U/t=1$      &Total energy ($/t$)& $(E_c/E_{c}^{\mathrm{APG}})*100$ ($\%$)  \\ \midrule
 Exact & -6.601158 & -  \\
 APG     & -6.596032 &  100 \\
 Rank-6 APG& -6.574143 &  77.2064  \\
 Rank-5 APG& -6.574142 &  77.2054  \\
 Rank-4 APG& -6.572637 &  75.6385 \\
 Rank-3 APG& -6.571406 &  74.3571  \\
 Rank-2 (a) APG& -6.570309 & 73.2143   \\
 Rank-2 (b) APG& -6.546232 &  48.1430  \\
 Rank-2 (c) APG& -6.525388 &   26.4373 \\
 Rank-1 APG& -6.500000 &  0  \\
 HF   & -6.500000 &   0 \\ \bottomrule
\end{tabular}}
\label{low-rank_x2}
\end{table}

\begin{figure}
\centering
{\resizebox*{10cm}{!}{\includegraphics{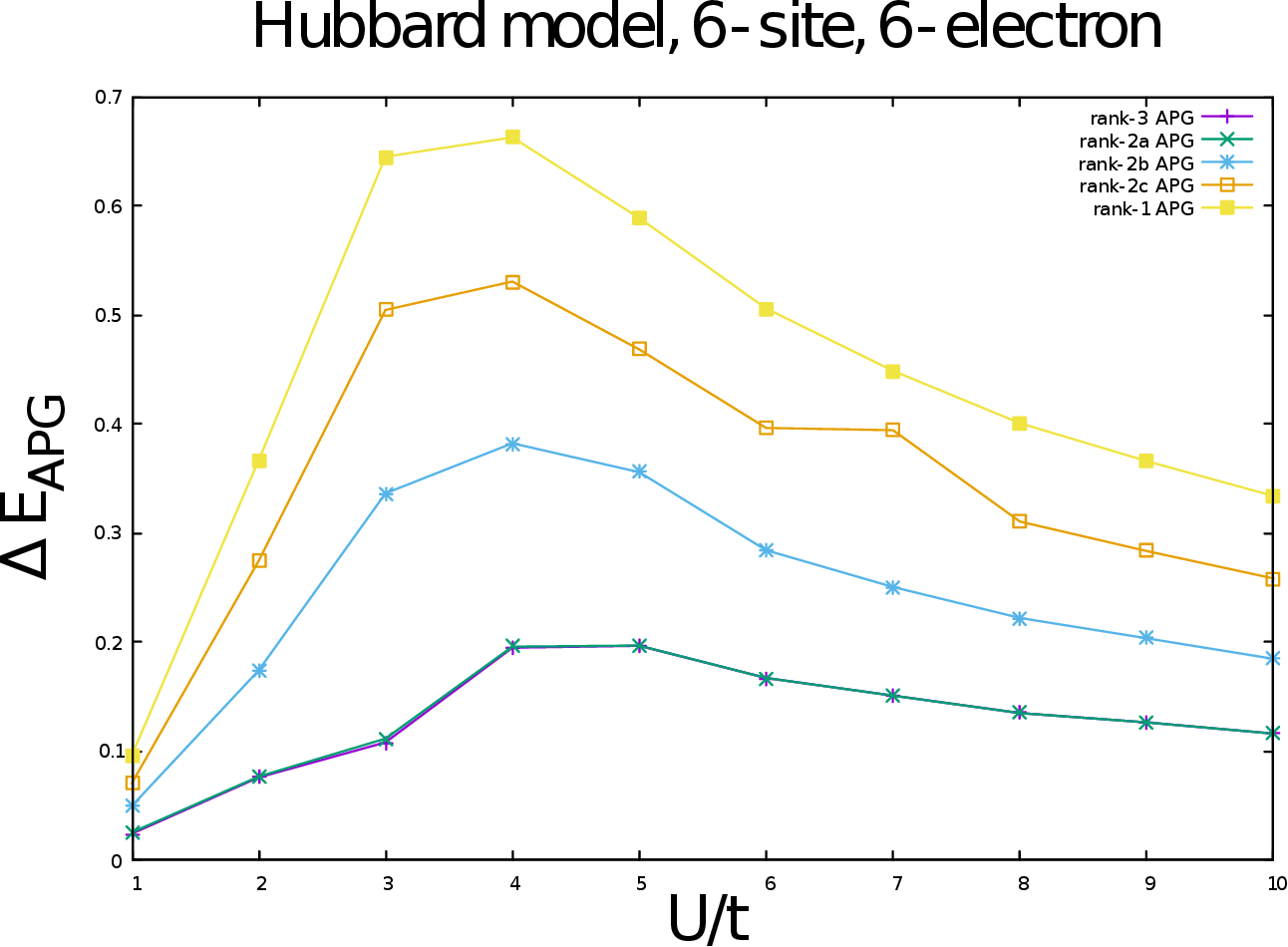}}}
\caption{Residual error of the total energy of rank-1 APG[$\lambda, X$] $\sim$ rank-3 APG[$\lambda, X$] in the Hubbard model ($U/t=1\sim10$ and half-filling) plotted against $U/t$.} 
\label{Udependency}
\end{figure}

These results showed that the accuracy improves as the rank increases.
It should be noted that the rank-6 APG, the full-rank APG in case of 6-site system, does not reproduce the original APG.
This is because we introduced approximation that makes all unitary transformations being identical.
The accuracy is almost the same for rank-2 (a) and ranks after rank-3, confirming that the original APG can be reproduced at a rank number of $N/2=3$ in a 6-electron system if there is no unitary approximation.
Figure \ref{Udependency} showed the results when changing the value of $U/t$ in the Hubbard model (6-site and 6-electron system). Even if $U/t$ was changed, rank-2 (a) and rank-3 APG had the same accuracy.

\begin{figure}
\centering
\subfloat[Rank-2 (a) up]{%
\resizebox*{4.5cm}{!}{\includegraphics{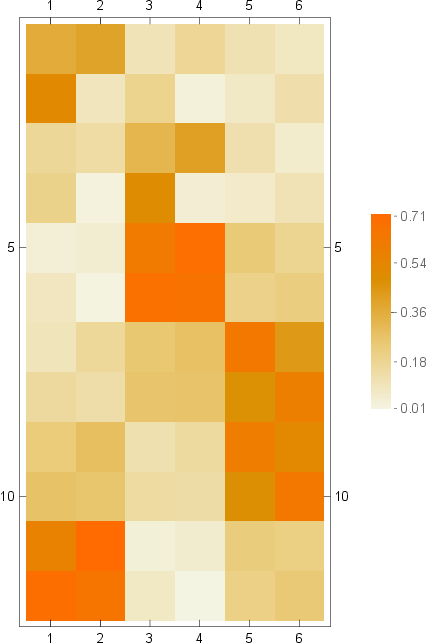}}}\hspace{5pt}
\subfloat[Rank-2 (b) up]{%
\resizebox*{4.5cm}{!}{\includegraphics{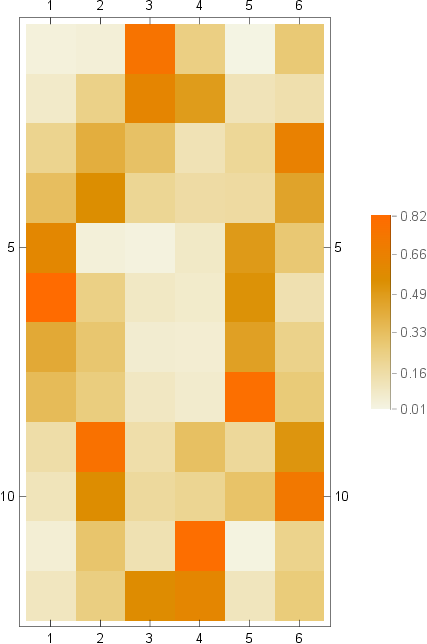}}}\hspace{5pt} 
\subfloat[Rank-2 (c) up]{%
\resizebox*{4.5cm}{!}{\includegraphics{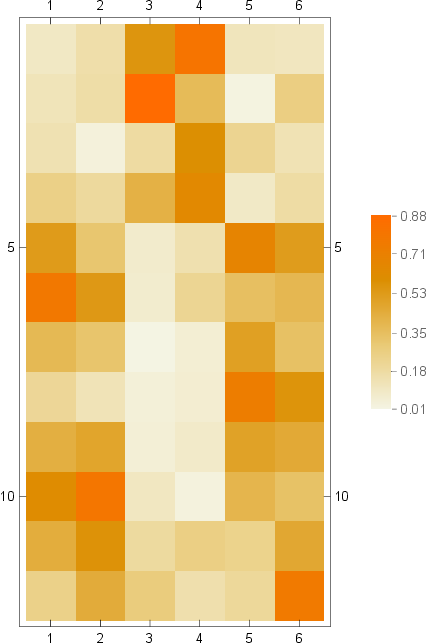}}} \vspace{10pt} \\
\subfloat[Rank-2 (a) down]{%
\resizebox*{4.5cm}{!}{\includegraphics{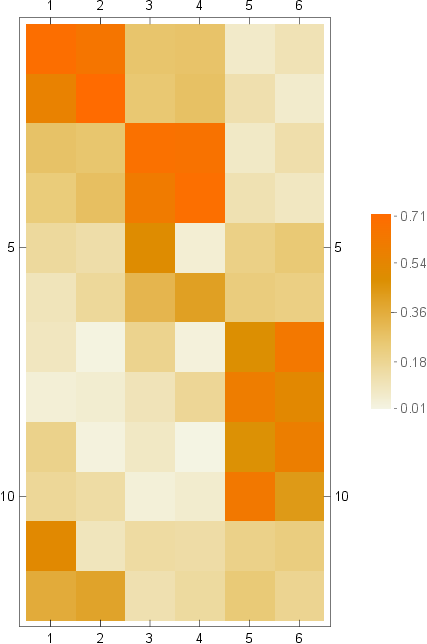}}}\hspace{5pt}
\subfloat[Rank-2 (b) down]{%
\resizebox*{4.5cm}{!}{\includegraphics{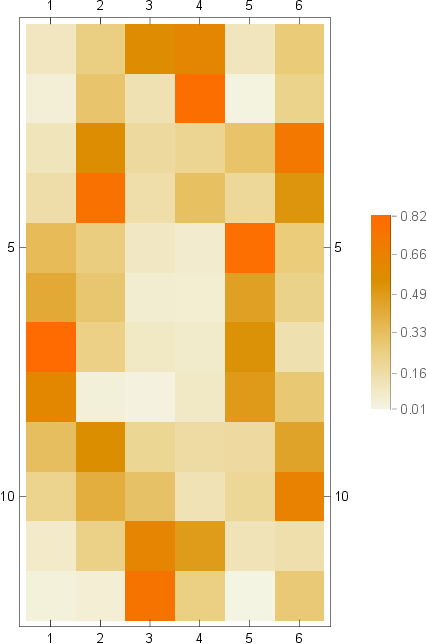}}}\hspace{5pt}
\subfloat[Rank-2 (c) down]{%
\resizebox*{4.5cm}{!}{\includegraphics{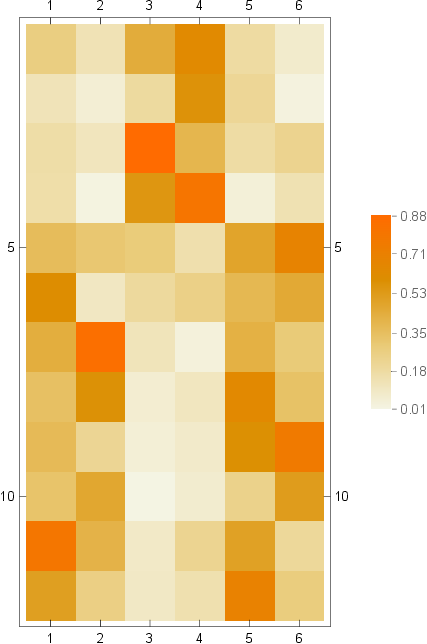}}}
\caption{Absolute values of the elements of the unitary matrices after optimization in rank-2 (a), (b), (c) APG[$\lambda, X$]. Captions (a), (b), and (c) are for the up site, and (d), (e), and (f) are for the down site.} 
\label{r2unitary}
\end{figure}

From the results of rank-2 APG, it was also found that changing the form of the wavefunction, even for the same rank, can make a difference in accuracy.
This is probably because each geminal matrix in the rank-2 (a) APG is strongly orthogonalized, and as a result, the rank-2 (a) wavefunction is the same as the generalized valence bond perfect-pairing (GVB-pp) wavefunction \cite{bobrowicz1977methods}.
Figure \ref{r2unitary} showed the absolute values of the elements of the unitary matrices of rank-2 (a), (b), and (c) after orbital optimizations. In rank-2 (a), the element values are similar between the 1-2, 3-4, and 5-6 sites, indicating that bond orbital pairs are formed.

Furthermore, we calculated the first-order density matrices of the low-rank APG's as shown below.
\begin{eqnarray}
\rho_i = \bra{\Psi}\hat{c}^\dag_i\hat{c}_i \ket{\Psi}
\end{eqnarray}

\begin{figure}
\centering
{%
\resizebox*{7cm}{!}{\includegraphics{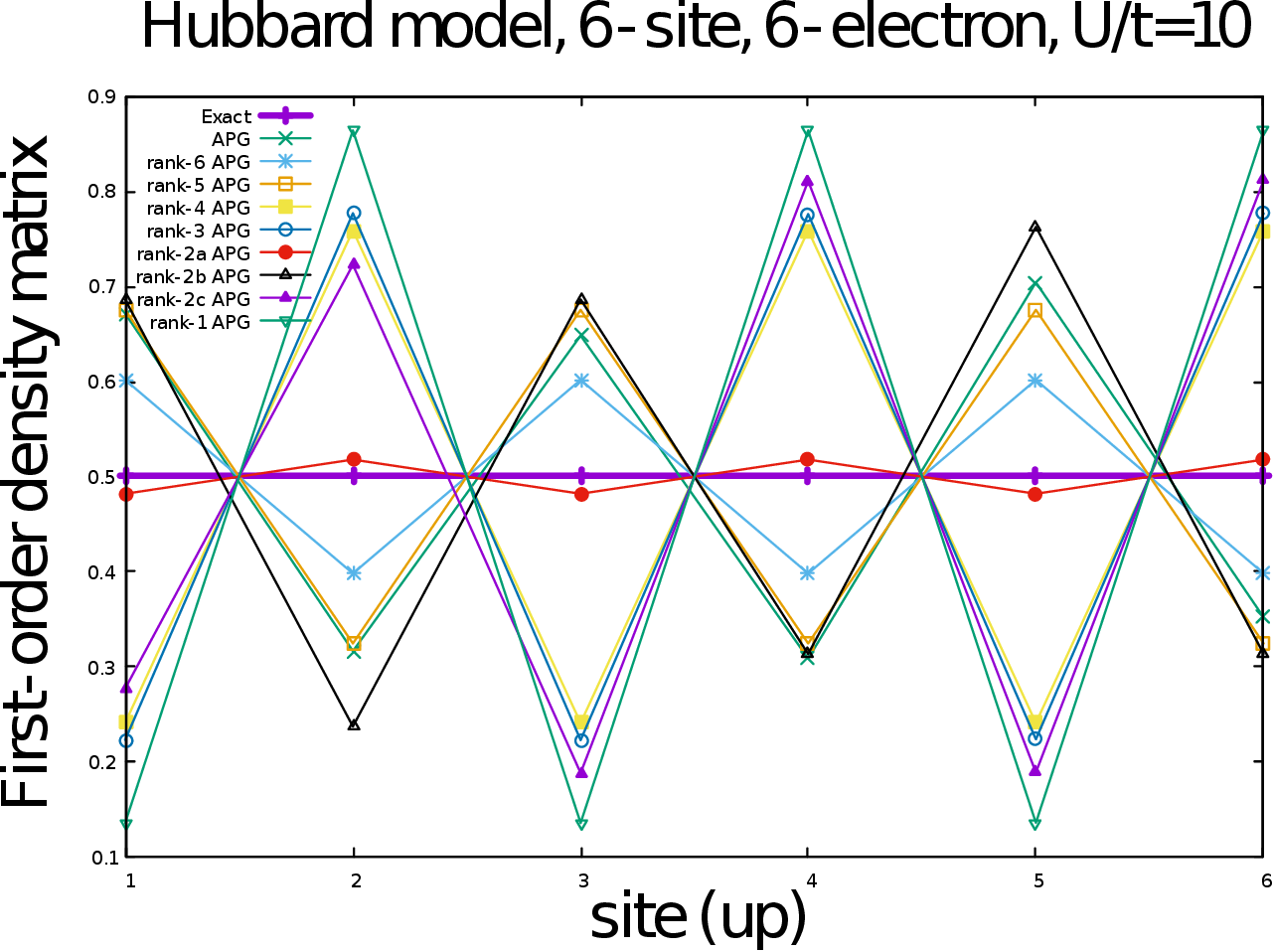}}}
{%
\resizebox*{7cm}{!}{\includegraphics{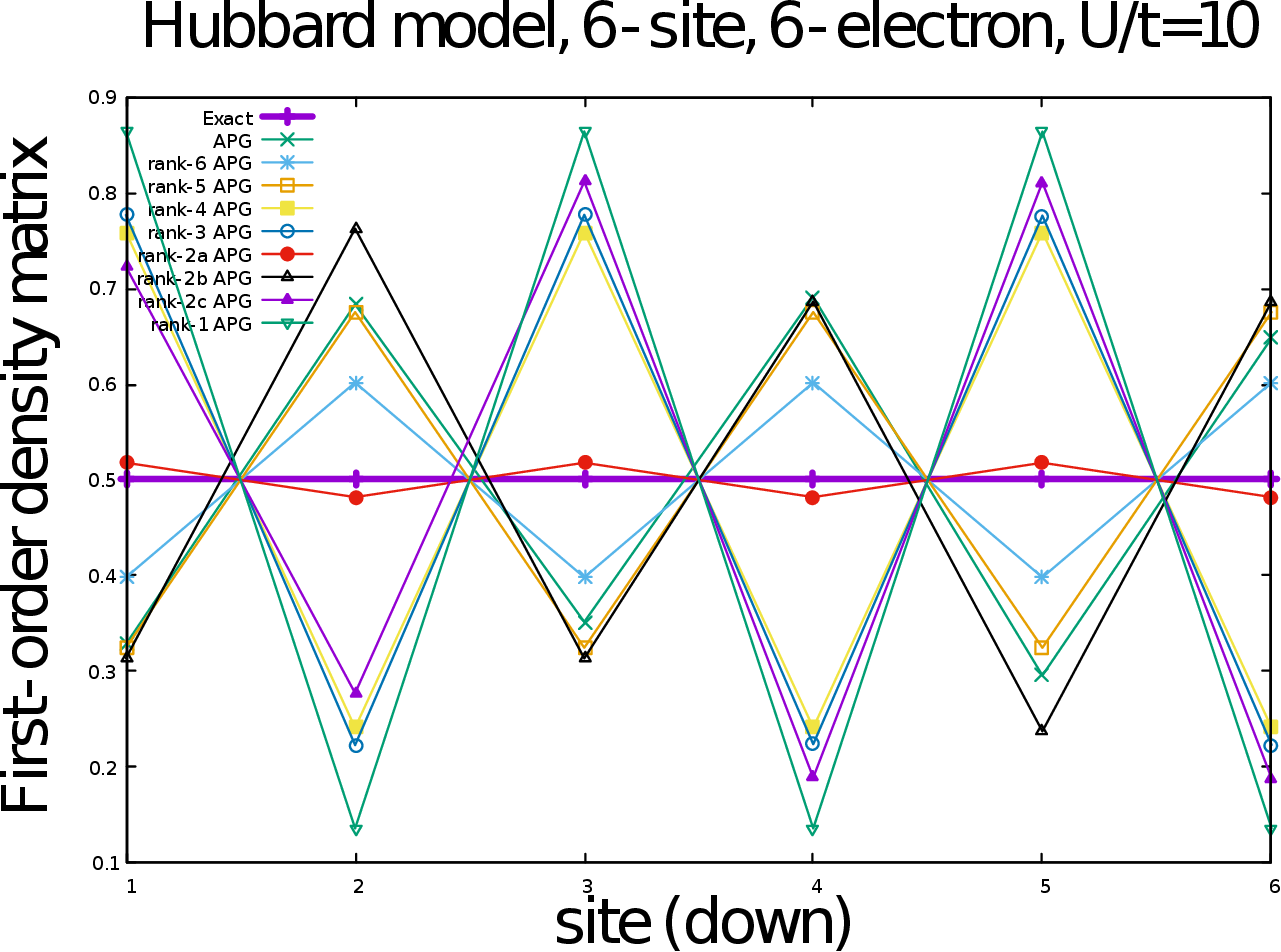}}}
\caption{The first-order density matrices of the exact, APG and low-rank APG[$\lambda, X$] in the Hubbard model ($U/t=10$, 6-site and 6-electron system).} 
\label{dm1}
\end{figure}

Figure \ref{dm1} showed the first-order density matrices for the exact, the APG, and the low-rank APG[$\lambda, X$] in the Hubbard models ($U/t=10$, 6-site and 6-electron system).
From these results, we concluded that the rank-2 (a) can accurately represent the wavefunction.

Also, table \ref{low-rank_x3} showed the total energies for the low-rank APG in H$_{2}$O molecule with STO-6G basis sets.
Here we used the following wavefunctions.
\begin{eqnarray}
\ket{\Psi_{\mathrm{r1APG}}} &=& \lambda[1]^{(1)} \hat{a}^{\dag}_{1} \hat{a}^{\dag}_{\bar{1}} \lambda[2]^{(2)} \hat{a}^{\dag}_{2} \hat{a}^{\dag}_{\bar{2}} \lambda[3]^{(3)} \hat{a}^{\dag}_{3} \hat{a}^{\dag}_{\bar{3}} \lambda[4]^{(4)} \hat{a}^{\dag}_{4} \hat{a}^{\dag}_{\bar{4}} \lambda[5]^{(5)} \hat{a}^{\dag}_{5} \hat{a}^{\dag}_{\bar{5}}  \ket{0} \\
\ket{\Psi_{\mathrm{r2APG}}} &=& \left(\lambda[1]^{(1)} \hat{a}^{\dag}_{1} \hat{a}^{\dag}_{\bar{1}} +\lambda[1]^{(6)} \hat{a}^{\dag}_{6} \hat{a}^{\dag}_{\bar{6}}  \right)\left(\lambda[2]^{(2)} \hat{a}^{\dag}_{2} \hat{a}^{\dag}_{\bar{2}} +\lambda[2]^{(7)} \hat{a}^{\dag}_{7} \hat{a}^{\dag}_{\bar{7}}  \right)\left(\lambda[3]^{(3)} \hat{a}^{\dag}_{3} \hat{a}^{\dag}_{\bar{3}} +\lambda[3]^{(1)} \hat{a}^{\dag}_{1} \hat{a}^{\dag}_{\bar{1}}  \right)  \nonumber \\
&&\left(\lambda[4]^{(4)} \hat{a}^{\dag}_{4} \hat{a}^{\dag}_{\bar{4}} +\lambda[4]^{(2)} \hat{a}^{\dag}_{2} \hat{a}^{\dag}_{\bar{2}}  \right)  
\left(\lambda[5]^{(5)} \hat{a}^{\dag}_{5} \hat{a}^{\dag}_{\bar{5}} +\lambda[5]^{(3)} \hat{a}^{\dag}_{3} \hat{a}^{\dag}_{\bar{3}}  \right) \ket{0} \\
\ket{\Psi_{\mathrm{r3APG}}} &=& \left(\lambda[1]^{(1)} \hat{a}^{\dag}_{1} \hat{a}^{\dag}_{\bar{1}} +\lambda[1]^{(6)} \hat{a}^{\dag}_{6} \hat{a}^{\dag}_{\bar{6}}+\lambda[1]^{(4)} \hat{a}^{\dag}_{4} \hat{a}^{\dag}_{\bar{4}}   \right)
\left(\lambda[2]^{(2)} \hat{a}^{\dag}_{2} \hat{a}^{\dag}_{\bar{2}} +\lambda[2]^{(7)} \hat{a}^{\dag}_{7} \hat{a}^{\dag}_{\bar{7}} +\lambda[2]^{(5)} \hat{a}^{\dag}_{5} \hat{a}^{\dag}_{\bar{5}}  \right) \nonumber  \\
&&\left(\lambda[3]^{(3)} \hat{a}^{\dag}_{3} \hat{a}^{\dag}_{\bar{3}} +\lambda[3]^{(1)} \hat{a}^{\dag}_{1} \hat{a}^{\dag}_{\bar{1}} +\lambda[3]^{(6)} \hat{a}^{\dag}_{6} \hat{a}^{\dag}_{\bar{6}}  \right)  
\left(\lambda[4]^{(4)} \hat{a}^{\dag}_{4} \hat{a}^{\dag}_{\bar{4}} +\lambda[4]^{(2)} \hat{a}^{\dag}_{2} \hat{a}^{\dag}_{\bar{2}} +\lambda[4]^{(7)} \hat{a}^{\dag}_{7} \hat{a}^{\dag}_{\bar{7}}   \right)  \nonumber \\
&&\left(\lambda[5]^{(5)} \hat{a}^{\dag}_{5} \hat{a}^{\dag}_{\bar{5}} +\lambda[5]^{(3)} \hat{a}^{\dag}_{3} \hat{a}^{\dag}_{\bar{3}} +\lambda[5]^{(1)} \hat{a}^{\dag}_{1} \hat{a}^{\dag}_{\bar{1}}  \right) \ket{0}
\end{eqnarray}
Though we were only able to obtain results up to rank-3 because the orbital optimization took a very long time due to bad convergence character, the APG energy can be reproduced to a reasonable extent even at rank-3.

\begin{table}
\tbl{Total energy and the percentage of correlation energy ($E_{c}=E-E^{\mathrm{HF}}$) of exact diagonalization, APG, low-rank APG[$\lambda, X$] and HF in H$_{2}$O (STO-6G).}
{\begin{tabular}{lcc} \toprule
 H$_{2}$O   &Total energy (Hartree) & $(E_c/E_{c}^{\mathrm{APG}})*100$ ($\%$)  \\ \midrule
 Exact & -75.728706 &  - \\
 APG   & -75.728417 &   100  \\
 Rank-3 APG&  -75.721747 &  86.5876  \\
 Rank-2 APG&  -75.720566 &  84.2128  \\
 Rank-1 APG&  -75.678686 & 0   \\
 HF     & -75.678686 & 0  \\ \bottomrule
\end{tabular}}
\label{low-rank_x3}
\end{table}

\subsection{Low-rank AGP}

We also calculated the low-rank AGP to investigate the accuracy in the absence of approximations that treat unitary transformations as identical.
Table \ref{low-rank_agp_hubbard} showed the total energies for the low-rank AGP in the Hubbard model ($U/t=10$, 6-site and 6-electron system).
It can be seen that the accuracy improves as the rank increases, and rank-6 AGP is able to reproduce the original AGP.

\begin{table}
\tbl{Total energy and the percentage of correlation energy ($E_{c}=E-E^{\mathrm{HF}}$) of exact diagonalization, AGP, low-rank AGPs and HF in the Hubbard model ($U/t=10$, 6-site and 6-electron system).}
{\begin{tabular}{lcc} \toprule
  $U/t=10$      &Total energy ($/t$)& $(E_c/E_{c}^{\mathrm{AGP}})*100$ ($\%$)  \\ \midrule
 Exact & -1.664363 &   -\\
 AGP     & -1.263868&  100 \\
 Rank-6 AGP& -1.263865 &  99.9968 \\
 Rank-5 AGP& -1.263828 &  99.9473 \\
 Rank-4 AGP& -1.251625 &  83.8109  \\
 Rank-3 AGP& -1.188243 &   0 \\
 HF   & -1.188243 &  0  \\ \bottomrule
\end{tabular}}
\label{low-rank_agp_hubbard}
\end{table}

\section{Conclusions}

In this paper, we focused on the APG wavefunction, which is a representative geminal theory, and explored how accurately the geminal theory can describe electron correlation. Applying the geminal theories to the Hubbard model and small molecules, we found that the APG has better accuracy compared to AGP-CI, AGP and HF.
Further analysis of the geminal wavefunction reveals that the number of important eigenvalues in the geminal matrix is up-to $N/2$. Therefore, we developed the low-rank APG theory that extracts only the important eigenvalues and reconstructs the wavefunction, and confirmed how well the APG can be reproduced. 

As a result of applying low-rank APG[$\lambda, X$] to the Hubbard model and H$_{2}$O, almost the same accuracy was obtained for rank-2 (a) and ranks after rank-3. Although it is not as accurate as the original APG due to the unitary approximation, it was confirmed that $N/2$ ranks are almost as accurate as full rank. It was also found that the accuracy changes even for the same rank by changing the form of the wavefunction. Rank-2 (a) APG was found to be accurate in total energy and density matrix, and to accurately represent the wavefunction.

By using low-rank AGP, we also confirmed that in the absence of unitary approximation, low-rank AGP with increased ranks can reproduce the original AGP.

The low-rank APG[$\lambda, X$] proposed in this paper is difficult to optimize, and at this stage it is not a method that can be calculated at low computational cost. However, the low-rank APG results revealed that the APG wavefunction contains a lot of unnecessary information to express the electron correlation. Therefore, advancing the development of such methods will lead to low-cost and highly accurate calculation methods. Low-rank APG can also adjust how much electron correlation is incorporated as needed. This suggests the importance of developing a physically meaningful approximation theory that follows the essence of electron correlation.

\section*{Acknowledgements}

Some calculations were performed at the Research Center for Computer Science (RCCS), Institute for Molecular Science (IMS), Okazaki, Japan (Project: 24-IMS-C125).

\section*{Disclosure statement}

The authors report there are no competing interests to declare.

\section*{Funding}

This work was financially supported by JSPS KAKENHI (Grant Number 21H01893).

\bibliography{cite}

%\begin{thebibliography}{99}

%\end{thebibliography}{99}

\appendix

\section{Total energy formula for the AGP wavefunction} \label{app1}

The matrix elements of Hamiltonian of AGP wavefunction and AGP-CI wavefunction are analytically given by Onishi Yoshida formula as follows.
\begin{eqnarray}
\bra{F[{\lambda}]}c^{\dag}_{a}c_{b}\ket{F[\mu]} = \left(\frac{F[{\mu}]F[{\lambda}]^{\dag}t^{2}}{1+F[{\mu}]F[{\lambda}]^{\dag}t^{2}}\right)_{ba} \left.\exp\left(  \frac{1}{2}\mathrm{tr}\left[  \ln(1+F[{\mu}]F[{\lambda}]^{\dag}t^{2})\right] \right)\right| _{t^{N}} 
\end{eqnarray}
\begin{eqnarray}
\bra{F[{\lambda}]}c^{\dag}_{p}c^{\dag}_{q}c_{s}c_{r}\ket{F[{\mu}]} &=& \left( \left[\frac{F[{\mu}]F[{\lambda}]^{\dag}t^{2}}{1+F[{\mu}]F[{\lambda}]^{\dag}t^{2}}\right]_{rp}\left[\frac{F[{\mu}]F[{\lambda}]^{\dag}t^{2}}{1+F[{\mu}]F[{\lambda}]^{\dag}t^{2}}\right]_{sq} \right.\nonumber \\
&& -\left[\frac{F[{\mu}]F[{\lambda}]^{\dag}t^{2}}{1+F[{\mu}]F[{\lambda}]^{\dag}t^{2}}\right]_{rq}\left[\frac{F[{\mu}]F[{\lambda}]^{\dag}t^{2}}{1+F[{\mu}]F[{\lambda}]^{\dag}t^{2}}\right]_{sp} \nonumber\\
&& \left. +\left[\frac{t}{1+F[{\mu}]F[{\lambda}]^{\dag}t^{2}}F[{\mu}]\right]_{rs}\left[F[{\lambda}]^{\dag}\frac{t}{1+F[{\mu}]F[{\lambda}]^{\dag}t^{2}}\right]_{qp} \right)\nonumber\\
&&\times\left.\exp\left(  \frac{1}{2}\mathrm{tr}\left[  \ln(1+F[{\mu}]F[{\lambda}]^{\dag}t^{2})\right] \right)\right| _{t^{n}}
\end{eqnarray}
Also, since the derivative can be transformed as follows, it can be obtained analytically as well.
\begin{eqnarray}
\frac{\partial}{\partial F[{\mu}]_{cd}}\braket{F[{\lambda}]|F[{\mu}]} = \bra{F[{\lambda}]}c^{\dag}_{c}c^{\dag}_{d}\ket{F[{\mu}]} 
\end{eqnarray}

\section{Low-rank APG[$u$]} \label{app2}

Another form of the low-rank APG approximation is directly parameterized by independent without operator conversion.
For example, the rank-1 geminal is as follows.
\begin{eqnarray}
\hat{F}[k] &=& \sum_{ab}^{2K}F[k]_{ab} \hat{c}^{\dag}_{a}\hat{c}^{\dag}_{b} \nonumber \\
&\rightarrow& \sum_{ab}^{2K}\frac{\lambda[k]^{(p)}}{2}\left(U[k]_{ap}U^{*}[k]_{\bar{p}b} - U[k]_{a\bar{p}}U^{*}[k]_{pb} \right) \hat{c}^{\dag}_{a}\hat{c}^{\dag}_{b} \nonumber  \\
&\equiv& \sum_{ab}^{2K}\left(u[k]_{ap}u^{*}[k]_{\bar{p}b} - u[k]_{a\bar{p}}u^{*}[k]_{pb} \right) \hat{c}^{\dag}_{a}\hat{c}^{\dag}_{b} \nonumber  \\
& \equiv& \hat{u}[k]_p \label{rank-1u}
\end{eqnarray}
In the third line of Eq (\ref{rank-1u}), the eigenvalues $\lambda$ and the unitary matrix $U$ are combined and replaced by $u$, and the last line defines the rank-1 geminal operator as $\hat{u}$ where $p$ represents an orbital.

For example, when $N/2=3$, the rank-1 APG wavefunction can be written as follows.
\begin{eqnarray}
\ket{\Psi_{\mathrm{r1APG}} }&=& \hat{u}[1]_1 \hat{u}[2]_2 \hat{u}[3]_3 \ket{0} \nonumber \\
&=& \frac{1}{6} \left(\hat{u}[1]_1 +\hat{u}[2]_2 +\hat{u}[3]_3 \right)^3 \ket{0} \label{rank-1eg}
\end{eqnarray}
The second line of Eq (\ref{rank-1eg}) is derived using the fact that $\hat{u}$ is rank-1, for example $\hat{u}[1]_{1}\hat{u}[1]_{1}=0$. The following holds for the general number of electrons.
\begin{eqnarray}
\hat{u}[1]_{1}\cdots \hat{u}[n]_{n} \ket{0}= \frac{1}{n!} \left(  \hat{u}[1]_{1} +\cdots + \hat{u}[n]_{n} \right)^{n} \ket{0}
\end{eqnarray}
Note that the rank-1 APG is equivalent to HF. This holds true for both low-rank APG methods. See Appendix \ref{app3} for details. 

The Hamiltonian contains one-body and two-body terms, and we aim to rewrite the matrix elements of the Hamiltonian into the same form as the overlap of the AGP-CI wavefunction.
In order to make the one-body term act on the wavefunction,
\begin{eqnarray}
\sum_{pq} t_{pq}\hat{c}^{\dag}_{p}\hat{c}_{q} \ket{\Psi} \equiv \hat{t}\ket{\Psi}
\end{eqnarray}
we first consider acting on the rank-1 operator $\hat{f}[1]=\sum_{ab}f[1]_{ab}\hat{c}^{\dag}_{a}\hat{c}^{\dag}_{b}$,  ($= \hat{u}[1]_{1}$).
\begin{eqnarray}
\hat{t} \hat{f}[1] &=& \sum_{pq}t_{pq}\hat{c}^{\dag}_{p}\hat{c}_{q} \sum_{ab}f[1]_{ab}\hat{c}^{\dag}_{a}\hat{c}^{\dag}_{b} \nonumber \\
&=& 2\sum_{ab} (tf[1])_{ab} \hat{c}^{\dag}_{a}\hat{c}^{\dag}_{b} +\hat{f}[1] \hat{t} \nonumber \\
&\equiv & \hat{t}[1] + \hat{f}[1] \hat{t} 
\end{eqnarray}
Therefore, when it acts on the wavefunction, it becomes as follows.
\begin{eqnarray}
\hat{t}\ket{\Psi} &=& \hat{t} \hat{f}[1] \cdots \hat{f}[n] \ket{0} \nonumber \\
&=& \left( \hat{t}[1] + \hat{f}[1] \hat{t}  \right) \hat{f}[2] \cdots \hat{f}[n] \ket{0} \nonumber \\
&=& \hat{t}[1]\hat{f}[2] \cdots \hat{f}[n] \ket{0}  + \cdots + \hat{f}[1] \cdots \hat{f}[n-1] \hat{t}[n] \ket{0} 
\end{eqnarray}
Since $\hat{t}[1], \cdots, \hat{t}[n]$ are the rank-2 operators (for example, $\hat{t}[1]^3=0$), the following holds.
\begin{eqnarray}
\hat{t}[1]\hat{f}[2] \cdots \hat{f}[n] \ket{0}  &=& \frac{1}{2n!} \left[ \left(  \hat{t}[1] + \hat{f}[2] + \cdots + \hat{f}[n]  \right)^{n} - \left( - \hat{t}[1] + \hat{f}[2] + \cdots + \hat{f}[n]  \right)^{n}  \right] \nonumber \\
\end{eqnarray}
Now we have the AGP-CI form, so Eq (\ref{OYovl}) can be applied.

When calculating the two-body term, the AGP-CI form is obtained as follows.
\begin{eqnarray}
\hat{c}^{\dag}_{p} \hat{c}^{\dag}_{q} \hat{f}[1] \cdots \hat{f}[n] &\equiv& \hat{v}  \hat{f}[1] \cdots \hat{f}[n] \nonumber \\
&=& (\hat{v} + \hat{f}[1] +  \cdots + \hat{f}[n]  )^{n+1}
\end{eqnarray}
Here $\hat{v}=\sum_{kl}v_{kl}\hat{c}^{\dag}_{k} \hat{c}^{\dag}_{l} $ is a rank-1 operator and $v$ becomes
\begin{eqnarray}
v_{kl} = \left\{ \begin{array}{ll}
0.5 & (k,l) = (p, q) \\
-0.5 & (k,l) = (q, p) \\
0 & (\mathrm{otherwise}) 
\end{array} \right.  .
\end{eqnarray}

When calculating the derivative required for the variational calculation, first consider differentiating $f$ with respect to $u$.
\begin{eqnarray}
\frac{\partial f_{pq}}{\partial u_{ak}} &=& \delta_{ap}u_{q\bar{k}}-\delta_{aq}u_{p\bar{k}} \\
\frac{\partial f_{pq}}{\partial u_{a\bar{k}}}& =& - \delta_{ap}u_{qk}+\delta_{aq}u_{pk} 
\end{eqnarray}
Therefore, differentiating the wavefunction with respect to $u$ gives as follows.
\begin{eqnarray}
\frac{\partial }{\partial u_{ak}} \ket{\Psi} &=& \left(\sum_{pq} \left(\delta_{ap}u_{q\bar{k}}-\delta_{aq}u_{p\bar{k}} \right)\hat{c}^{\dag}_{p}\hat{c}^{\dag}_{q} \right) \prod_{1\leq l \leq n }^{l\neq k} \hat{f}[l]\ket{0} \nonumber \\
&=& \frac{1}{n!}\left(  \left(\sum_{pq} \left(\delta_{ap}u_{q\bar{k}}-\delta_{aq}u_{p\bar{k}} \right)\hat{c}^{\dag}_{p}\hat{c}^{\dag}_{q} \right)+  \sum_{1\leq l \leq n } ^{l\neq k}\hat{f}[l]\right)^n \ket{0}\\
\frac{\partial }{\partial u_{a\bar{k}}} \ket{\Psi} &=& \left(\sum_{pq} \left(-\delta_{ap}u_{qk}+\delta_{aq}u_{pk} \right)\hat{c}^{\dag}_{p}\hat{c}^{\dag}_{q} \right)  \prod_{1\leq l \leq n }^{l\neq k} \hat{f}[l]\ket{0} \nonumber \\
&=& \frac{1}{n!}\left(  \left(\sum_{pq} \left(-\delta_{ap}u_{qk}+\delta_{aq}u_{pk} \right)\hat{c}^{\dag}_{p}\hat{c}^{\dag}_{q} \right)+  \sum_{1\leq l \leq n } ^{l\neq k}\hat{f}[l]\right)^n \ket{0}
\end{eqnarray}
The differential term can also be transformed into AGP-CI form, and all elements required for variational calculation can be analytically calculated using Eq (\ref{OYovl}).

The rank-2 APG takes two eigenvalues of each geminal matrix to construct the wavefunction. For example, when $N/2 = 3$, it becomes
\begin{eqnarray}
\ket{\Psi_{\mathrm{r2APG}} } =\left(\hat{u}[1]_1 +\hat{u}[1]_2\right)\left(\hat{u}[2]_3+\hat{u}[2]_4\right)\left( \hat{u}[3]_5+\hat{u}[3]_6 \right)  \ket{0} .
\end{eqnarray}
The rank-2 APG wavefunction can be calculated in the same way as rank-1 because it becomes a linear combination of rank-1 APGs when expanded. In general, when the rank-$m$ APG is expanded, $m^{N/2}$ terms appear. However, since the number of terms appeared by decomposing APG using Fischer's formula is $2^{N/2-1}$, computing in a manner similar to APG is less computationally expensive than expanding to rank-1 form.

The rank-$m$ APG in the $N$-electron system becomes
\begin{eqnarray}
\ket{\Psi_{\mathrm{r}m\mathrm{APG}} } = \left(\hat{u}[1]_1 +\cdots +\hat{u}[1]_m\right)\cdots \left( \hat{u}[N/2]_{m(N/2-1)+1}+\cdots+\hat{u}[N/2]_{mN/2} \right) \ket{0} . \nonumber \\ 
\end{eqnarray}

Note that when optimizing with this low-rank APG[$u$] method, we do not impose any restrictions on $u$. Since the matrix $U$ that combines the vectors $u$ is a unitary matrix, it is necessary to ensure unitarity. Therefore, in order for the unitary matrix $U$ derived from $u$ to have unitary property, we add the following term to forcibly ensure unitary property during optimization.
\begin{eqnarray}
E=\frac{\langle \Psi|\hat{H}|\Psi\rangle }{\langle \Psi|\Psi\rangle} + \sum_k\left(\|\left( U[k]U[k]^{T} - I \right)\|_{F}\right)^2*10000 
\end{eqnarray}
Here $\| A\|_{F} = \sqrt{\sum_i\sum_jA_{ij}^2}$ is the Frobenius norm. 
The derivative of the product of unitary matrices can also be calculated as follows.
\begin{eqnarray}
 \frac{\partial } {\partial u[1]_{ak}} U[1]U[1]^{T}  = \left( \frac{\partial } {\partial u[1]_{ak}} U[1]\right) U[1]^{T} + U[1] \left(\frac{\partial } {\partial u[1]_{ak}} U[1]^{T} \right)
\end{eqnarray}

Table \ref{low-rank_u} shows the total energy results for low-rank APG[$u$] in the Hubbard model ($U/t=10$, 6-site and 6-electron system). The rank-4 was less accurate than the rank-3. Also, overall accuracy is not very good. This is thought to be due to the optimization performed by forcibly ensuring unitary property. 

\begin{table}
\tbl{Total energy and the percentage of correlation energy ($E_{c}=E-E^{\mathrm{HF}}$) of exact diagonalization, APG, low-rank APG[$u$] and HF in the Hubbard model ($U/t=10$, 6-site and 6-electron system).}
{\begin{tabular}{lcc} \toprule
 $U/t=10$  &Total energy ($/t$)& $(E_c/E_{c}^{\mathrm{APG}})*100$ ($\%$)  \\ \midrule
 Exact & -1.66436 &  - \\
 APG   & -1.52224 &   100 \\ 
 Rank-4 APG& -1.311475 &  36.8964  \\
 Rank-3 APG& -1.387966 &  59.7983  \\
 Rank-2 APG& -1.318732 &  39.0693 \\
 Rank-1 APG& -1.18824 &  0 \\
 HF     & -1.18824 & 0  \\ \bottomrule
\end{tabular}}
\label{low-rank_u}
\end{table}

\section{HF wavefunction} \label{app3}

\begin{figure}
\centering
{\resizebox*{7cm}{!}{\includegraphics{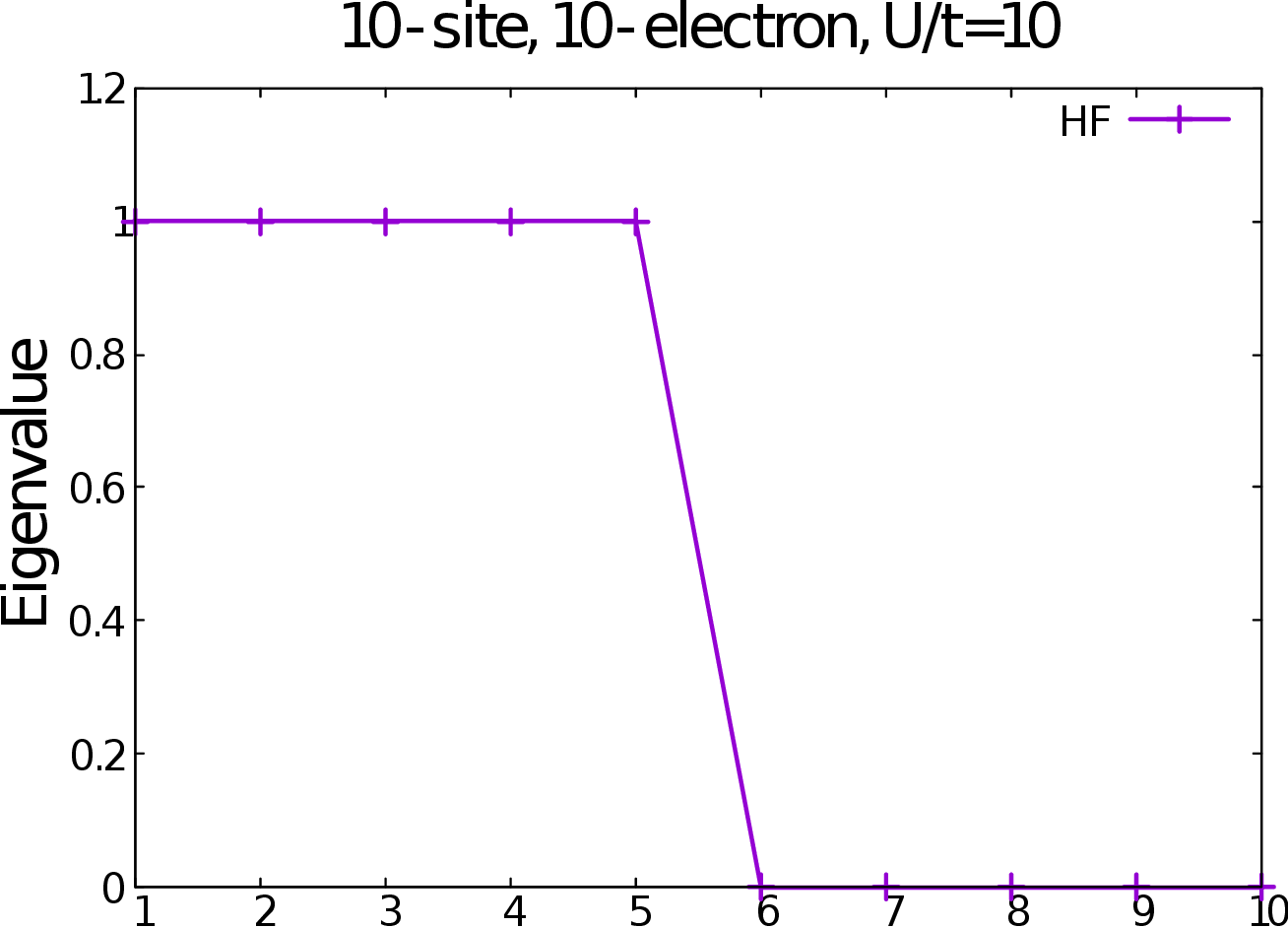}}}
\caption{The absolute eigenvalues of geminal matrices in HF in the Hubbard model (10-site, 10-electron and $U/t=10$).} 
\label{hf_eigen}
\end{figure}

The HF wavefunction is defined as
\begin{eqnarray}
\ket{\Psi_{\mathrm{HF}}}&=&  \prod^N_{i}\left(\sum_a \Phi_{ai}c^{\dag}_a\right)\ket{0}  \nonumber \\
&=& \prod^N_{i} \psi^{\dag}_i\ket{0} .
\end{eqnarray}
Here $\Phi$ is a normalized orthogonal basis as
\begin{eqnarray}
\sum_{a} \Phi^{*}_{ai}\Phi_{aj} = \delta_{ij} .
\end{eqnarray}
Then, the HF becomes
\begin{eqnarray}
\ket{\Psi_{\mathrm{HF}}} &=& \prod^{N/2}_{i} \psi^{\dag}_{2i-1} \psi^{\dag}_{2i}\ket{0}  \nonumber \\
&=& \left( \sum^{N/2}_i \psi^{\dag}_{2i-1} \psi^{\dag}_{2i} \right)^{\frac{N}{2}}\ket{0}  \nonumber \\
&=& \left( \sum^{N/2}_i \left(\sum_a \Phi_{a,2i-1}c^{\dag}_a\right) \left(\sum_b \Phi_{b,2i}c^{\dag}_b\right) \right)^{\frac{N}{2}}\ket{0}  \nonumber \\
&=& \left( \sum_{a,b} \left( \sum^{N/2}_i\Phi_{a,2i-1}\Phi_{b,2i} \right) c^{\dag}_ac^{\dag}_b\right)^{\frac{N}{2}}\ket{0} .
\end{eqnarray}
Therefore, one can represent $F$ as
\begin{eqnarray}
F_{ab}=  \sum^{N/2}_i\left(\Phi_{a,2i-1}\Phi_{b,2i} - \Phi_{b,2i-1}\Phi_{a,2i} \right)  \label{HFgeminal}.
\end{eqnarray}
Comparing Eq (\ref{HFgeminal}) and Eq (\ref{rank-1u}), it is found that the HF wavefunction corresponds to the case of rank-1 APG. Figure \ref{hf_eigen} shows the result of calculating the eigenvalues similar to Figure \ref{hubbard_eigen} after expressing HF using a geminal matrix as in Eq (\ref{HFgeminal}). In HF, $N/2$ eigenvalues have a value of 1 and the $N/2$ eigenvalues have a value of 0.

\section{Computational details} \label{app4}

We applied the APG and other geminal wavefunctions to the Hubbard model as shown below and small molecules (hydrogen chain (H$_6$), Li$_{2}$ and H$_{2}$O). 
\begin{eqnarray}
H=t\sum_{<ij>}\hat{c}^\dag_i\hat{c}_j+U\sum_{(ij)}\hat{c}^\dag_i\hat{c}^\dag_j\hat{c}_i\hat{c}_j
\end{eqnarray}
where $t$ is the hopping integral, $U$ is the on-site Coulomb interaction, the element $<i,j>$ is the nearest-neighbor pair of the same spin and $(i, j)$ is the on-site pair. Also, periodic boundary conditions were used. For the small molecules, STO-6G was used as the basis function. The molecular coordinates are [H(0, 0, 0), H(0, 0, 1.40218), H(0, 0, 2.80435), H(0, 0, 4.20653), H(0, 0, 5.60871), H(0, 0, 7.01088)] in H$_6$, [Li(0, 0, 0), Li(0, 0, 5.0512)] in Li$_{2}$, and [O(0, 0, 0), H(-1.809, 0, 0), H(0.453549, 1.751221, 0)] in H$_{2}$O (all units are in Bohr). Integrals of molecular Hamiltonian were calculated using PySCF. We use the variational calculation to optimize the trial wavefunction and get the ground state energy. Optimizations were performed using the conjugate gradient method, and using all the elements of $F$, $\lambda$, $X$ and $u$ as variational variables. The results were compared with the exact diagonalization results using the H$\Phi$ package \cite{KAWAMURA2017180} and HF results using the mVMC package \cite{misawa2018mvmc}. In all results, the number of geminal types in AGP-CI is the same as in APG. For example, in a six-electron system, APG uses three types of geminals, so APG-CI also uses three types of geminals.

\end{document}